\documentclass[amsmath,amssymb,aps,pre,showpacs,twocolumn,superscriptaddress]{revtex4-2}
\usepackage[dvipsnames,table,xcdraw]{xcolor}
\usepackage[utf8]{inputenc}
\usepackage{hyperref}
\usepackage{wrapfig}
\usepackage{bm}
\usepackage[capitalise]{cleveref}
\usepackage{float}
\usepackage{graphicx,epstopdf,subfigure}

\usepackage{hyperref}
\usepackage{amsmath}  
\usepackage{amssymb}
\usepackage{gensymb}
\usepackage{enumitem}
\usepackage{indentfirst}
\usepackage{nicematrix}
\usepackage{phaistos}

\colorlet{R}{violet!80}
\colorlet{T}{teal!80}
\colorlet{O}{olive!80}
\colorlet{V}{violet!60}
\colorlet{Y}{Dandelion}
\colorlet{E}{Emerald}
\colorlet{P}{Thistle!60}
\colorlet{M}{MidnightBlue!80}
\colorlet{B}{BlueViolet!80}

\begin{abstract}
A common approach for analyzing hypergraphs is to consider the projected adjacency or Laplacian matrices for each order of interactions (e.g., dyadic, triadic, etc.). However, this method can lose information about the hypergraph structure and is not universally applicable for studying dynamical processes on hypergraphs, which we demonstrate through the framework of cluster synchronization. Specifically, we show that the projected network does not always correspond to a unique hypergraph structure. This means the projection does not always properly predict the true dynamics unfolding on the hypergraph. Additionally, we show that the symmetry group consisting of permutations that preserve the hypergraph structure can be distinct from the symmetry group of its projected matrix. Thus, considering the full hypergraph is required for analyzing the most general types of dynamics on hypergraphs. We show that a formulation based on node clusters and the corresponding edge clusters induced by the node partitioning, enables the analysis of admissible patterns of cluster synchronization and their effective dynamics. Additionally, we show that the coupling matrix projections corresponding to each edge cluster synchronization pattern, and not just to each order of interactions, are necessary for understanding the structure of the Jacobian matrix and performing the linear stability calculations efficiently. 
\end{abstract}

\begin{document}
	\title{Analyzing states beyond full synchronization on hypergraphs requires methods beyond projected networks}
	\author{Anastasiya Salova}
	\email[]{avsalova@ucdavis.edu}
	\affiliation{Department of Physics and Astronomy, University of California, Davis, CA 95616, USA}
	\affiliation{Complexity Sciences Center, University of California, Davis, CA 95616, USA}
	\author{Raissa M. D'Souza}
	\affiliation{Complexity Sciences Center, University of California, Davis, CA 95616, USA}
	\affiliation{
		Department of Computer Science and
		Department of Mechanical and Aerospace Engineering, University of California, Davis, CA 95616, USA}
	\affiliation{Santa Fe Institute, Santa Fe, NM 87501, USA}
	
	\maketitle

\section{Introduction}

The framework of dynamical systems on dyadic networks provides a useful tool for modeling the behavior of many systems, including those from biological, social, and engineered realms \cite{barrat2008dynamical,danon2011networks,rhoden2012self,porter2020nonlinearity+,curto2019relating}. However, some systems have higher order non-additive interactions which require going beyond dyadic interactions \cite{bick2021higherorder,battiston2020networks}. 
Hypergraphs are a natural extension of dyadic networks that allow the study of a wider range of systems by capturing higher-order interactions. 
Naturally, adding higher order interactions requires modifying tools from systems with dyadic interactions to be applicable to dynamics on hypergraphs and also developing new tools to analyze the system's behavior. 

There are several ways higher order dynamics can be defined. Namely, dynamics can be defined on the nodes interacting via hyperedges of different orders \cite{skardal2019abrupt,skardal2020memory,gambuzza2020master,zhang2020unified,xu2020bifurcation,landry2020effect,lucas2020multi,skardal2020higher}. Alternatively, especially if the dynamics is defined on a simplicial complex, the dynamical signals can be defined on
simplices of different dimensions \cite{millan2020explosive,ghorbanchian2021higher,deville2021consensus}. Here, we take the former approach.
Specifically, we consider dynamics on undirected hypergraphs, where the evolution of each node depends on the state of its neighbors via dyadic and higher order interactions. Additionally, we assume that some sets of nodes within the system have similar internal dynamics,  and some sets of edges have similar coupling forms. We specifically study cluster synchronization where groups of oscillators in the system have fully synchronized trajectories, but distinct groups follow distinct trajectories. 
The framework of cluster synchronization is useful for analyzing intricate patterns of synchronization in dynamical systems on hypergraphs and it illustrates the difference between analysis based on full hypergraph considerations and those based on dyadic projections.

To study synchronization in higher order systems, the generalization of the dyadic graph adjacency and Laplacian matrices are useful tools.
Several ways to generalize these matrices from the perspective of node interactions have been recently proposed \cite{carletti2020dynamical,lucas2020multi,de2021phase, gambuzza2020master}.
These generalizations are based on projecting the higher order edges onto dyadic cliques and finding the adjacency or Laplacian of a resulting network for each order of interactions.
Specifically, the projections of this form are sufficient to formulate stability conditions for full synchronization on undirected hypergraphs \cite{gambuzza2020master, de2021phase, lucas2020multi} and chemical hypergraphs \cite{mulas2020coupled} or even some cases of cluster synchronization, such as non-intertwined cluster synchronization \cite{zhang2020unified}. A downside of hypergraph projection is the non-applicability of such analysis to more intricate types of synchronization dynamics in higher order systems. In this manuscript, we demonstrate that the hypergraph projection description is not sufficient for analyzing cluster synchronization in the most general case. 

First we show that the projection is not always in one-to-one correspondence with the original higher order system. In other words, several non-isomorphic hypergraphs can have the same projection onto a dyadic network. 
Specifically we demonstrate that distinct hypergraphs can have the same projection, yet the effective interactions on the hypergraphs can be distinct even for the same pattern of cluster synchronization (i.e., which nodes follow the same trajectory, and which do not). It is these effective interactions between the clusters that determine the dynamical behavior. Projections are sensitive enough for capturing full synchronization dynamics and its stability properties, but do not necessarily capture more intricate patterns of synchronization.

We next compare the symmetries of the full hypergraph with the symmetries of its dyadic projections to show that the hypergraph does not always admit the same cluster synchronization patterns as one would deduce from its dyadic projections. 
Symmetry considerations, namely the orbits of the symmetry group of the system (as well as its subgroups), can be used to determine some of the admissible cluster synchronization states \cite{golubitsky2003symmetry,pecora2014cluster}. While the symmetries of the projected hypergraph are often in direct correspondence with the symmetries of the original hypergraph \cite{gambuzza2020master}, we demonstrate that for some topologies, some of the symmetries of the projected network do not preserve the hypergraph structure (also discussed in Ref.\cite{mulas2020hypergraph}). 
 
Our final contribution is showing how projected networks can be used for stability calculations. In systems with purely dyadic interactions, cluster synchronization states do not necessarily arise from symmetries alone \cite{stewart2003symmetry}. 
They can also arise from more general balanced equivalence relations.
This is also the case for systems with higher order interactions, both for Laplacian-like coupling \cite{salova2021h1,salova2021code} and more general couplings discussed in this manuscript. 
To analyze general cluster synchronization patterns whether they arise from symmetries or more generally from equitable partitions, we define the concept of \textit{edge clusters} with each edge cluster corresponding to a specific \textit{edge synchronization pattern}. We demonstrate that one needs to define a separate projected adjacency matrix for each edge synchronization pattern and hyperedge order to fully capture the structure of the Jacobian matrix used for linear stability analysis.

Linear stability calculations can be simplified using simultaneous block diagonalization \cite{zhang2020symmetry}.
We demonstrate that the set of matrices that need to be simultaneously block diagonalized to analyze cluster synchronization on hypergraphs includes the projected adjacency matrices for each edge pattern of synchronization for interactions beyond dyadic (discussed in detail in \cref{subsec: app}). In contrast, stability analysis for dyadic interactions does not require tracking the individual edge synchronization patterns.

The rest of the manuscript is organized as follows. \cref{sec: background} provides the basic formulation for dynamical systems on undirected hypergraphs and the general conditions for cluster synchronization in such systems based on node and edge partitions. \cref{sec: projection} demonstrates that the hypergraph projection does not always allow us to unambiguously reconstruct the original hypergraph up to an isomorphism, which can produce misleading predictions for the effective dynamics of cluster synchronization states. In \cref{sec: symm} we consider symmetries and show that some of the orbital partitions of the projected hypergraph do not describe the admissible cluster synchronization states of the original hypergraph, thus projected adjacency matrices are not always sufficient to determine the admissible patterns of synchronization on hypergraphs. \cref{sec: stab} demonstrates that the projected hypergraph adjacency matrices combined with the cluster synchronization indicator matrices are not sufficient to fully represent the structure of the Jacobian and simplify its analysis in the case of the most general hypergraph structure and pattern of synchronization. Instead, we show how to use projections corresponding to different cluster synchronization patterns to perform the linear stability analysis. Finally, we discuss our results and future directions in \cref{sec: conclusion}. 

\section{Background: cluster synchronization oh hypergraphs} \label{sec: background}
\subsection{Hypergraph structure and dynamics}\label{subsec: structure and dyn}
First, we define the general form of the dynamics on hypergraphs that is being considered. 
A hypergraph is defined by a set of $N$ nodes and a set of hyperedges $e_j\in\mathcal{E}$. In this work, we focus on undirected hyperedges. 
Let $\mathcal{E}_i\subset \mathcal{E}$ be the set of hyperedges 
that contain node $i$. Each hyperedge $e_j\in\mathcal{E}_i$ contains a set of nodes $e_j=\{i,j_1,...,j_{m-1}\}$. The order of the hyperedge $e_j$
is $m$, which is the number of nodes including $i$ that are part of it. Thus, $m=2$ corresponds to dyadic edges, $m=3$ to triadic edges, etc.

Using notation similar to Ref.\cite{de2021phase}, we can express the evolution of the state of each node in the system, $x_i\in R^n$, as:
\begin{align}
	\dot{x}_i = F_i(x_i) + \sum\limits_{e\in\mathcal{E}_i} G_{e}(x_i,x_{e\backslash i}).
	\label{eq:hyper}
\end{align}
Here, the function $F_i(x_i)$ describes the evolution of uncoupled nodes, and the function $G_e(x_i,x_{e\backslash i})$ is a coupling function corresponding to the influence of the hyperedge $e$ 
on node $i$, where $x_i$ is the state of the node $i$ itself, and $x_{e\backslash i}$ is the state of the rest of the edge. This setup is general, including the case when the interaction hypergraph is a simplicial complex which has the additional requirement that each subset of nodes in the hyperedge forms a hyperedge of lower order. 

Often, some degree of homogeneity is present within the nodal dynamics, $F_i(x_i)$, of different nodes $i$ as well as in the coupling dynamics, $G_e(x_i,x_{e\backslash i})$. 
In that case, one can use the hypergraph structure to find nontrivial partitions 
into sets of nodes that can fully synchronize. In the simplest case, all the self-dynamics are characterized by the same function $F$
and the coupling dynamics of a given order $m$ are characterized by the same function $G^{(m)}$. 
In that case, it is sufficient to consider adjacency structures (e.g., adjacency tensors) with binary entries. 

The exact higher order adjacency structure can be defined in terms of the collection of $m$ incidence matrices $I^{(m)}$, one for each order $m$. 
Let $\mathcal{E}^{(m)}_i$ be the set of hyperedges of order $m$ containing the node $i$. Then, the nonzero elements of  the incidence matrix are $[ I^{(m)}]_{i,e}=1$ if $e\in\mathcal{E}^{(m)}_i$. Additionally, we assume undirected coupling, so $[I^{(m)}]_{i,e}=1$ for all $i\in e$.

With these simplifications, the dynamics of \cref{eq:hyper} can be expressed as:
\begin{align} \label{eq: dynamics}
	\dot{x}_i=&F(x_i) + \sum\limits_{m=2}^{d}\sigma^{(m)}\sum\limits_{e\in \mathcal{E}^{(m)}} [I^{(m)}]_{i,e} G^{(m)}(x_i,x_{e\backslash i}),
\end{align}
where due to undirected coupling we assume that the function
$G^{(m)}(x_i,x_{e\backslash i})$ is invariant under any reordering of nodes in $x_{e\backslash i}$.

In Ref.\cite{salova2021h1}, we cover the stability analysis in the case of Laplacian and Laplacian-like coupling. Here, we assume more general undirected coupling. In the case of undirected coupling, the presence of the hyperedge $\{i_1,...,i_k,...,i_m\}$ providing input to node $i_1$ via the coupling function $G^{(m)}$, s.t. $\dot{x}_{i_1}=...+G^{(m)}(x_{i_1},...,x_{i_k},...,x_{i_m})$, implies that hyperedge affects $x_{i_{k}}$ via the same coupling function, s.t. $\dot{x}_{i_k}=...+G^{(m)}(x_{i_k},...,x_{i_1},...,x_{i_m})$. Additionally, the coupling function responsible for providing input into node $x_{i_1}$ has to be invariant with respect to permutations of the elements corresponding to the nodes providing this input within a hyperedge, namely, $G^{(m)}(x_{i_1},...,x_{i_k},...,x_{i_{k'}},...)=G^{(m)}(x_{i_1},...,x_{i_{k'}},...,x_{i_{k}},...)$. For a concrete example of triadic coupling, consider the extension of the Kuramoto model to triadic interactions presented in Ref.\cite{PhysRevResearch.2.023281}, with $G^{(3)}(x_i,x_j,x_k)=K\sin(\theta_j+\theta_k-2\theta_i)$, where we set the coupling strengths $\sigma^{(3)}$ to be identical for all triadic edges. First, we note that $G^{(3)}(x_i,x_j,x_k)=G^{(3)}(x_i,x_k,x_j)$. In addition, for the coupling to be undirected, we require that $I_{i,e}=I_{j,e}=I_{k,e}$, where the edge $e$ consists of nodes $i$, $j$, and $k$.

\subsection{Bipartite representation of a hypergraph}
While incidence matrices are a useful and compact representation of the hypergraph structure, sometimes it is helpful to deal with square matrices instead. Thus, hypergraphs represented via a bipartite graph will be useful for much of the analysis herein. The adjacency matrix $\mathcal{M}$ of the bipartite representation of a hypergraph is of the form: 
\begin{align}\label{eq: m}
	\mathcal{M}=\begin{pmatrix}
		0_{N\times N} & I_{N\times M}\\
		I^T_{N\times N} & 0_{M\times M}
	\end{pmatrix},
\end{align}
where $N$ is the number of nodes in the hypergraph, $M$ is the number of edges, and $I$ is its incidence matrix.
While this matrix is less compact than the incidence matrix, this bipartite graph representation allows the use of standard dyadic interaction tools in analyzing systems with higher order interactions, as $\mathcal{M}$ is a square matrix. 

An important caveat here is that one needs to additionally take into account that the elements of $\mathcal{M}$ represent the relations between nodes and edges, and not simply the interactions between the nodes. This is discussed in more detail in \cref{sec: projection} in the context of hypergraph isomorphism and \cref{sec: symm} in relation to admissible patterns of cluster synchronization. 

\subsection{Dyadic projections of hypergraphs}\label{subsec: hyper proj}

A common way to analyze hypergraph structure and full synchronization dynamics is by using the projection of the hypergraph structure onto a dyadic coupling matrix for each order of interaction. Depending on the type of the coupling function, either an adjacency or Laplacian projection can be used. In several recent publications \cite{carletti2020dynamical,lucas2020multi,de2021phase, gambuzza2020master}, the projected matrices for each order of interactions are defined as:
\begin{align}\label{eq: proj}
	\mathcal{A}^{(m)}=I^{(m)}[I^{(m)}]^T-\mathcal D^{(m)},
\end{align}
where $[\mathcal D^{(m)}]_{ii}=\sum\limits_{j}I^{(m)}_{ij}$ and has zero off-diagonal elements. 

This projection is useful in analyzing, for instance, the stability of full synchronization in systems with higher order interactions, by either forming an aggregate projection matrix with different edge orders being assigned different weights \cite{de2021phase,carletti2020dynamical,lucas2020multi}, or considering projected Laplacians in case of noninvasive coupling \cite{gambuzza2020master}. However, in some cases, this projection loses information about the original hypergraph even for a given order of interactions (e.g., triadic), as discussed in \cref{sec: projection} and \cref{sec: symm}. Additionally, these projections are insufficient for cluster synchronization analysis, which is why we need to define such a projection for every \textit{edge pattern of synchronization}, as discussed in \cref{sec: stab}.

\subsection{Admissible patterns of cluster synchronization on hypergraphs}

While projection matrices are useful in analyzing full synchronization, collective behavior of coupled dynamical systems is more complicated when the nodes are not fully synchronized. Often, it is useful to analyze these behaviors using the framework of cluster synchronization, where the nodes in the same clusters $C_i$ are fully synchronized due to receiving the same dynamical input, but their behavior is distinct from all the other clusters $C_j$. Cluster synchronization can arise as a form of symmetry breaking in systems with identical nodes and edges (or hyperedges of the same order) that allow full synchronization. However, it can also be present in systems with multiple node and edge types, such as multilayer networks, in which full synchronization solutions are not admissible. 

Patterns of cluster synchronization have been extensively analyzed for systems with dyadic interactions \cite{belykh2008cluster,pecora2014cluster,pecora2017discovering,cho2017stable,salova2020decoupled}, with a few recent advances to higher order systems. Cluster synchronization of coupled map lattices on chemical hypergraphs was recently analyzed in Ref.\cite{bohle2021coupled}, but the setup is distinct from the general structure and dynamics considered in this manuscript. Stability of cluster synchronization in systems like the one in \cref{eq: dynamics} is analyzed in Ref.\cite{zhang2020unified}. However, the question of admissibility of different patterns is not discussed there, and the analysis is limited to non-intertwined clusters. Finally, cluster synchronization on hypergraphs is briefly discussed in Ref.\cite{gambuzza2020master}. However, the reference only discusses the patterns of synchronization arising from symmetries and does not discuss the ones arising from more general partitions (e.g., discussed in Ref.\cite{salova2021h1} and later herein). Additionally, the conditions for symmetry-based clusters in Ref.\cite{gambuzza2020master} may not be sufficient for general hypergraphs, and additional checks must be performed as discussed in detail in \cref{sec: symm}.

\begin{figure}[h]
	\includegraphics[scale=.4]{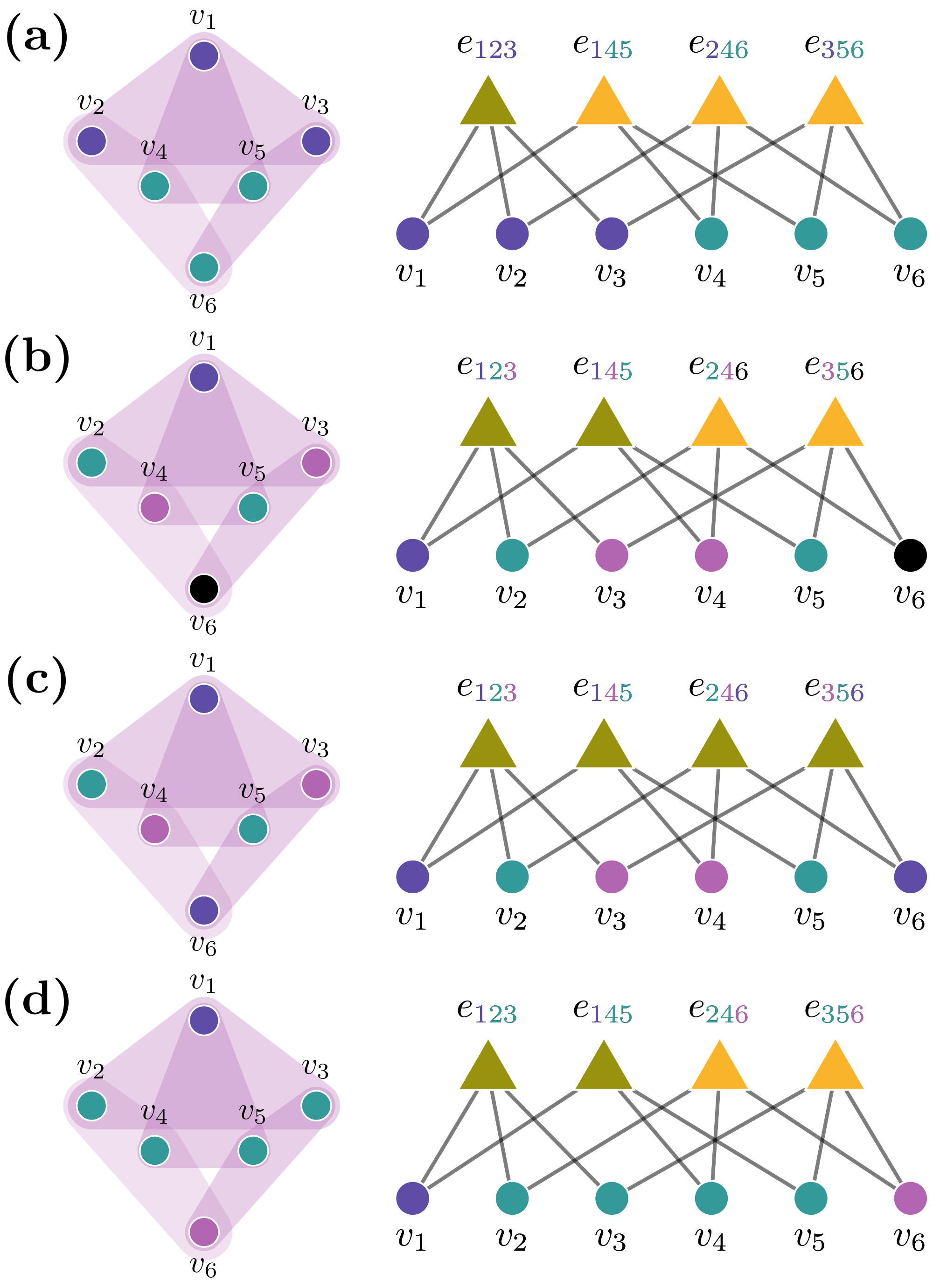}
	\caption{Synchronization patterns in hypergraphs. Left column: hypergraph, right column: equivalent bipartite representation. [(a-d)] Distinct cluster synchronization patterns.  
	}
	\label{fig: partitions}
\end{figure}

In this section, we demonstrate how to find the admissible cluster synchronization patterns by partitioning the nodes into node clusters, and the edges into edge clusters based on the node clusters those edges span. The framework is similar to that in Ref.\cite{salova2021h1}, but we do not restrict the coupling functions to \textit{Laplacian-like} coupling. As an example of cluster synchronization, consider \cref{fig: partitions}(a). The hypergraph structure shown on the left admits a cluster synchronization pattern with two node clusters, shown in purple and teal.
Each purple node $p$ obtains input from two hyperedges, one of the form $C^{(3)}_{ppp}$ (containing three nodes in the purple cluster) and one $C^{(3)}_{ptt}$. Each teal node $t$ gets input from two edges of the form $C^{(3)}_{ptt}$. We will index the node clusters as $C_j$ (where $j$ can refer to a cluster number or a cluster ``color''). Here, the node clusters are $C_1=C_p=\{1,2,3\}$ and $C_2=C_t=\{4,5,6\}$. The edge clusters induced by node partition (i.e., hyperedges which span equivalent node clusters, and, therefore, have equivalent node trajectories) are denoted by $C_j^{(m)}$. In this example, $C^{(3)}_1=C^{(3)}_{ppp}=\{[1,2,3]\}$ and $C^{(3)}_2=C^{(3)}_{ptt}=\{[1,4,5],[2,4,6],[3,5,6]\}$. The bipartite graph in \cref{fig: partitions}(a) (right) demonstrates the relations between nodes (circles) and edges (triangles). The bipartite graph makes it clear that there are two triadic edge clusters ($C^{(3)}_1$ shown in olive and $C^{(3)}_2$ shown in yellow) induced by the node clusters. 

The cluster synchronization pattern in \cref{fig: partitions}(a) is not the only admissible pattern. \cref{fig: partitions} (a-d) shows four distinct example partitions, using direct hypergraph representation (left column) and its bipartite representation (right column). 

Mathematically, the condition for an admissible cluster synchronization state based on the incidence matrix is
\begin{align}\label{eq: incidence cs}
	\sum\limits_{e_j\in C^{(m)}_k} I^{(m)}_{ij} = \sum\limits_{e_j\in C^{(m)}_k} I^{(m)}_{i'j}
\end{align}
where $i,i'\in C_l$, and the summation is performed over all the columns of $I^{(m)}$ corresponding to the edges in the $k$th edge cluster of order $m$, denoted by $C_k^{(m)}$. \cref{eq: incidence cs} has to hold for all the orders of interaction and edge clusters, unless the specific form of the coupling function makes some edge clusters irrelevant to cluster synchronization admissibility (e.g., fully synchronized hyperedges in Ref.\cite{salova2021h1}).

The effective interactions between different clusters are contained in the \textit{quotient hypergraph}, where 
\begin{align}
	I^{(m)}_{\text{eff}}=\mathcal{P}_{n}I^{(m)}\mathcal({\mathcal P}_{e}^{(m)})^T,
\end{align}
where $\mathcal{P}_n$ ($K\times N$) and $\mathcal{P}_e^{(m)}$ ($K_m\times N$) are the indicator matrices corresponding to node and edge partitions, and $I^{(m)}$ is the $m$th order incidence matrix.
The nonzero elements of the indicator matrices $\mathcal{P}_n$ and $\mathcal{P}_e^{(m)}$ are defined by
$[\mathcal{P}_n]_{i,j} = 1$ if node $i$ belongs to node cluster $C_j$, and $[\mathcal{P}_e^{(m)}]_{i,j} = 1$ if the $m$th order edge $i$ belongs to the $m$th order edge cluster $C^{(m)}_j$.

Note that \cref{eq: incidence cs} can be easily modified to handle the case where there are different types of nodes and hyperedges in the system. If distinct node types are present, only the ones within the same type are expected to fully synchronize. To put it in the form of \cref{eq: incidence cs}, we can form a trivial incidence matrix $I^{(1)}$, where $I^{(1)}_i=I^{(1)}_j$ if and only if the nodes $i$ and $j$ are of the same type, and add those incidence matrices to the set that needs to be tested in \cref{eq: incidence cs}. If distinct hyperedge types are present, \cref{eq: incidence cs} has to hold for each edge interaction order $m$ and for each edge type.

\begin{figure}[]
	\includegraphics[scale=.4]{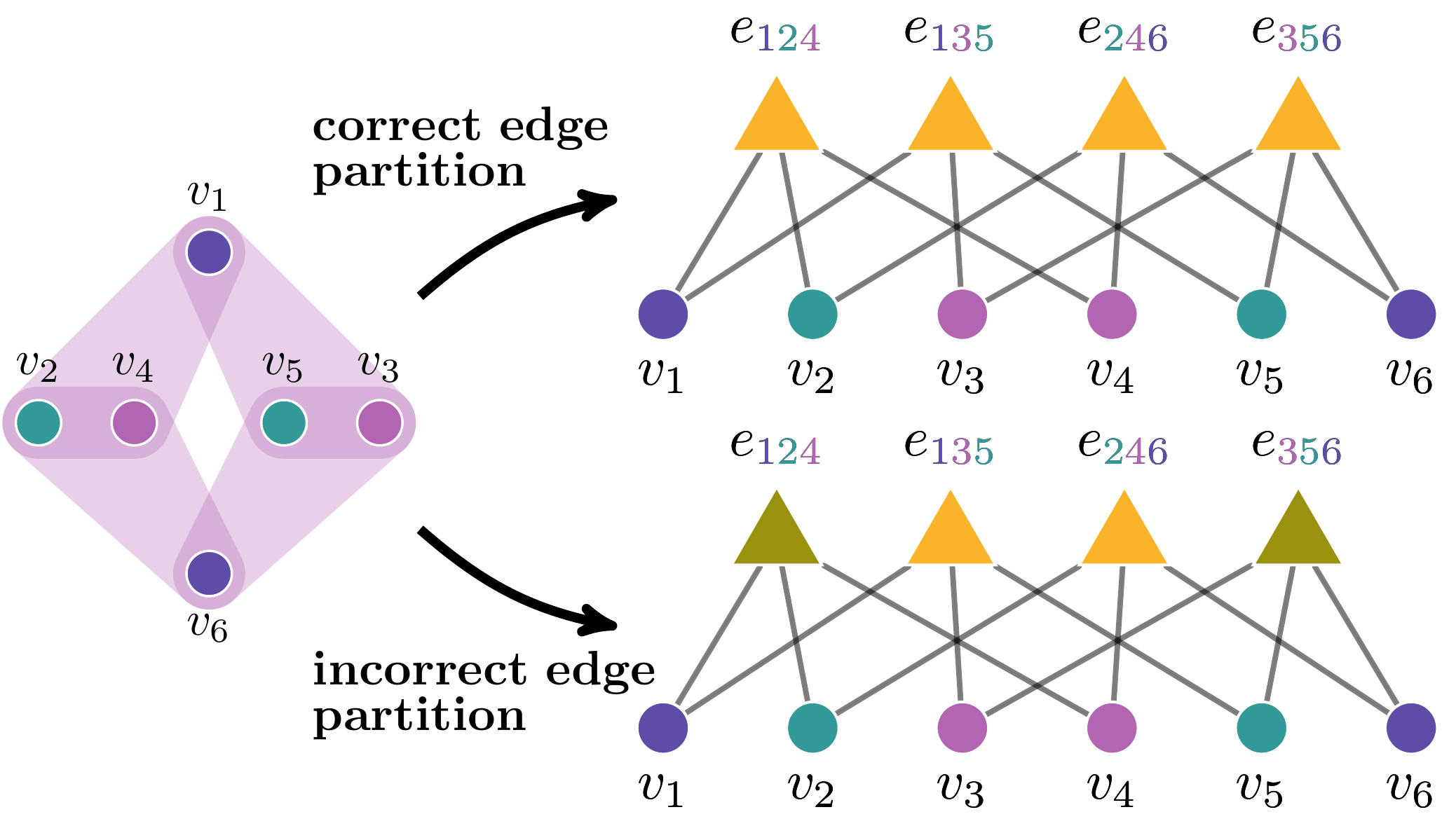}
	\caption{Left: cluster synchronization pattern on a hypergraph. Right: equitable partitions of the corresponding bipartite network. Only the top partition represents the correct node and edge partition of the hypergraph.
	In the bottom partition, the hyperedges $e_{124}$ and $e_{356}$ are incorrectly assigned into a cluster distinct from that containing $e_{135}$ and $e_{246}$, even though all the hyperedges consist of one violet, one purple, and one teal node.
	}
	\label{fig: partitions coarse}
\end{figure}

Equivalently, the bipartite graph adjacency matrix (\cref{eq: m}) can be used to partition nodes and edges into clusters using the methods applicable to systems with dyadic interactions (even for systems with different types of nodes and hyperedges). Importantly, since we distinguish between nodes and edges, they need to be partitioned into clusters separately (corresponding to the case of two distinct types of nodes in systems with dyadic interactions). It is also important to note that for each \textit{node} partition obtained from the bipartite representation, only the coarsest \textit{edge} partition is properly identified. For instance, \cref{fig: partitions coarse} demonstrates two partitions admissible on the bipartite graph, whose structure corresponds to the hypergraph with six nodes (shown as circles in the bipartite graph) and four triadic edges (shown as shaded triangles). However, only one of the resulting partitions (\cref{fig: partitions coarse} top right) is an admissible partition of the nodes and hyperedges of the hypergraph itself. \cref{fig: partitions coarse} bottom right shows that partitioning the bipartite representation can misidentify the edge partitions induced by the node partitions.
In this case, all hyperedges contain the same nodes (purple, teal, violet), and thus have to belong to the same edge cluster, although the bipartite representation divides them into two edge clusters.

\section{Hypergraph dyadic projection: loss of information on structure and effective dynamics}\label{sec: projection}

Hypergraph projections can be used to analyze fully synchronized states and their stability \cite{de2021phase,gambuzza2020master}. However, this tool is not always useful in analyzing general dynamics on hypergraphs, including cluster synchronization. Initial results obtained in Ref.\cite{de2021phase} led its authors to conjecture that it is possible to create a hypergraph projection (with the adjacency matrix calculated as a weighted sum of terms defined in \cref{eq: proj}) that fully preserves the information about the hypergraph structure. However, we show the projection as defined in \cref{eq: proj} does not necessarily correspond to a unique hypergraph. In fact, these distinct hypergraphs that get mapped onto the same single projection do not even have to be isomorphic, as we show next in \cref{subsec: structure}. As a result, sometimes the hypergraphs with the same node clusters and projected adjacency matrix have distinct quotient hypergraphs, and thus different cluster synchronization dynamics.

\subsection{Example: non-isomorphic hypergraphs with distinct effective dynamics but identical dyadic projection} \label{subsec: structure}
Identical hypergraphs, as well as isomorphic hypergraphs, produce identical dynamical behavior, including cluster synchronization. However, the hypergraphs that map onto the same projected network are not necessarily identical or isomorphic, and therefore can produce distinct dynamical behavior despite having the same projection. This can be investigated computationally, especially since the problem can be considered on a single order of higher order interactions at the time, because distinct orders can be distinguished in the projection.

To obtain hypergraphs that are not isomorphic, but which have the same projected dyadic adjacency matrix, it is sufficient to find two distinct incidence matrices, $I_1^{(m)}$ and $I_2^{(m)}$, that satisfy
\begin{align}\label{eq: distinct i}
	I_1^{(m)}[{I}_1^{(m)}]^T-D^{(m)}={I}_2^{(m)}[{I}_2^{(m)}]^T-D^{(m)}=\mathcal{A}^{(m)},
\end{align} 
with no nontrivial permutational matrix $P$ satisfying
\begin{align}\label{eq: no perm}
P(\mathcal{M}_1+\mathcal{R})=(\mathcal{M}_2+\mathcal{R})P,
\end{align}
where $\mathcal{M}_1$ and $\mathcal{M}_2$ are the respective adjacency matrices corresponding to the bipartite graph representation of the original hypergraphs. Here, $\mathcal{R}$ is a diagonal matrix whose purpose is to avoid permuting nodes with edges. It has diagonal entries $\mathcal{R}_{ii}=\alpha$ if $i\leq N$ and $\mathcal{R}_{ii}=\beta$ if $i> N$. In numerical calculations, $\alpha$ and $\beta$ can be set to be distinct random numbers.

\begin{figure*}
	\includegraphics[scale=.35]{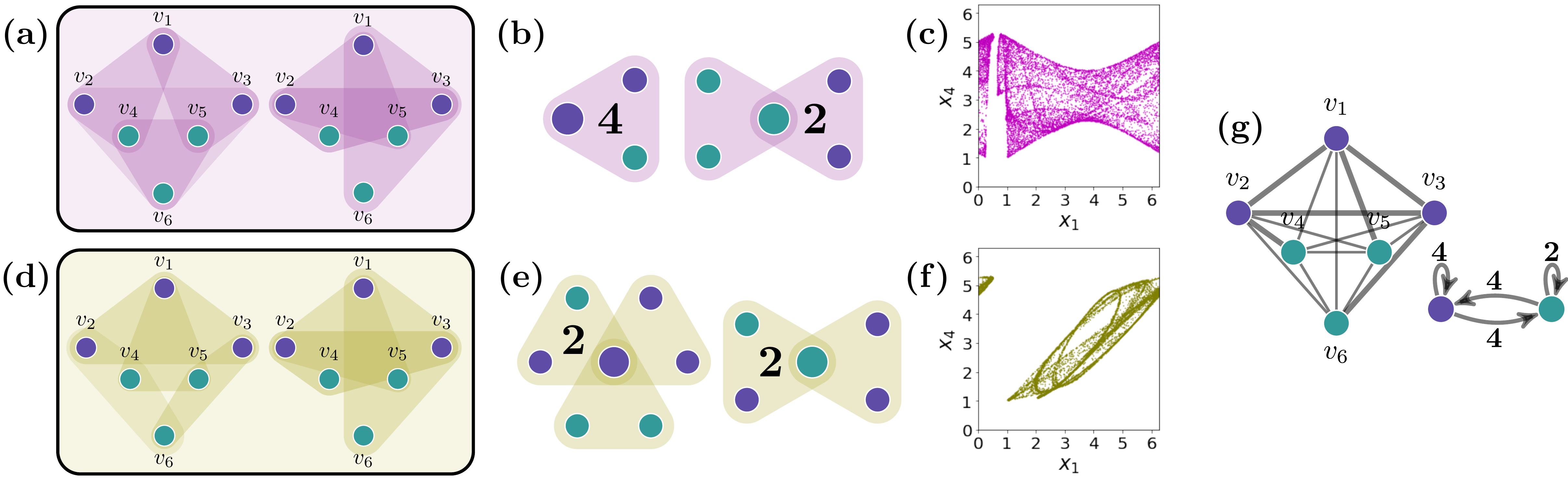}
	\caption{Synchronization patterns in hypergraphs. Teal and violet node colors correspond to distinct node synchronization clusters. [(a) and (d)] Hypergraph structure of two distinct hypergraphs (violet box (a) and olive box (d)) containing the union of hypergraphs shown on the left and on the right. The left hypergraphs in (a) and (d) are distinct, but isomorphic to each other. The right hypergraphs in these boxes are fully identical.  [(b) and (e)] Distinct quotient hypergraphs for cases (a) and (d) respectively. 
	[(c) and (f)] State of node $4$ ($x_4$) vs state of node $1$ ($x_1$) using \cref{eq: cluster dyn} dynamics evolved for $10^4$ time steps for the two-cluster state on hypergraphs (a) and (d) respectively. Node $1$ belongs to the violet cluster, node $4$ belongs to the teal cluster.
	(g) The same projected network (left) and its quotient network (right) results for both cases (a) and (d). Thick lines correspond to edges of weight two, and thin lines correspond to those of weight one.  
	}
	\label{fig: clusters}
\end{figure*}

As an example, consider two distinct hypergraphs with triadic interactions, each with six nodes and seven hyperedges. The first is shown in the box shaded in purple in \cref{fig: clusters}(a), and the second in the box shaded in olive in \cref{fig: clusters}(d). Both \cref{fig: clusters}(a) and \cref{fig: clusters}(d) are the union of the two simpler hypergraphs shown in the respective boxes. The hypergraphs in the left column of \cref{fig: clusters}(a) and (d) are isomorphic but distinct, whereas the right column hypergraphs are identical. Note, the full hypergraphs in \cref{fig: clusters}(a) and (d) are not isomorphic. 
The graph isomorphism problem is notoriously complicated. However, we used the \textit{networkx.is\_isomorphic} Python package \cite{hagberg2008exploring} to verify that indeed $\mathcal M_1+\mathcal{R}$ corresponding to \cref{fig: clusters}(a) is not isomorphic to $\mathcal M_2+\mathcal{R}$ corresponding to \cref{fig: clusters}(d). 
We denote their incidence matrices by $I_1$ and $I_2$.
The corresponding projected dyadic graph for both of the above hypergraphs, containing edges of weight $1$ and $2$ (shown in thin and thick lines respectively), is demonstrated in \cref{fig: clusters}(g). 
Its adjacency matrix is
\begin{align}
	\mathcal A^{(3)}&=I_1^{(3)}(I_{1}^{(3)})^T-\mathcal D^{(3)}=I_2^{(3)}(I_{2}^{(3)})^T-\mathcal D^{(3)}\nonumber \\&=
	\begin{pmatrix}
		0&1&1&2&2&2\\
		1&0&1&2&1&1\\
		1&1&0&1&2&1\\
		2&2&1&0&2&1\\
		2&1&2&2&0&1\\
		2&1&1&1&1&0\\
	\end{pmatrix}.
\end{align} 
Thus, the conditions from \cref{eq: distinct i} hold: two non-isomorphic hypergraphs, \cref{fig: clusters}(a) and (d), s.t. no nontrivial permutation satisfies \cref{eq: no perm} for their bipartite adjacency matrices $\mathcal{M}_1$ and $\mathcal{M}_2$, have the same projected adjacency matrix.

The fact that non-isomorphic hypergraphs may have the same projected graph has consequences on the cluster synchronization dynamics. Specifically, the inability to reconstruct the original hypergraph may lead to an ambiguity in effective dynamics in a hypergraph, even if the node assignment into clusters is the same between the two hypergraphs. As an example, consider the coloring of nodes on \cref{fig: clusters}. In all its subfigures, teal and violet nodes represent distinct clusters. This cluster assignment is admissible in both hypergraphs in \cref{fig: clusters}(a) and (d). The corresponding quotient hypergraphs are shown respectively in \cref{fig: clusters}(b) and (e). These hypergraphs represent the effective dynamics of each type of node (teal and violet). As very clearly visible in \cref{fig: clusters}, these quotient hypergraphs are qualitatively different. The dynamics on each type of nodes in case of \cref{fig: clusters}(b) is:
\begin{align}\label{eq: eff1}
	\dot x_p&=F(x_p)+4G^{(3)}(x_p,x_p,x_t),\nonumber\\
	\dot
	x_t&=F(x_t)+G^{(3)}(x_t,x_t,x_t)+2G^{(3)}(x_t,x_p,x_p),
\end{align}
whereas in case of \cref{fig: clusters}(e) it is:
\begin{align} \label{eq: eff2}
	\dot x_p=&F(x_p)+2G^{(3)}(x_p,x_p,x_t)
	\nonumber\\&+G^{(3)}(x_p,x_p,x_p)
	+G(x_p,x_t,x_t),\nonumber\\
	\dot
	x_t=&F(x_t)+2G^{(3)}(x_t,x_p,x_t)+G^{(3)}(x_t,x_p,x_p),
\end{align}
leading to distinct behaviors.

To provide a concrete example of distinct trajectories arising from \cref{eq: eff1} and \cref{eq: eff2}, we consider the discrete time dynamics:
\begin{align}	\label{eq: disc dyn}
	x_i^{t+1}=F(x_i^t)+\sigma^{(3)}\sum\limits_{e\in \mathcal{E}^{(m)}} [I^{(3)}]_{i,e} G^{(3)}(x_{e\backslash i}^t),
\end{align} 
where $x_i^t$ is the state of the node $i$ at time $t$, and the self evolution and coupling functions are defined as:
\begin{align}	\label{eq: cluster dyn}
	F(x_i^t)&=\alpha\dfrac{1-\cos(x_i^{t})}{2}+\dfrac{\pi}{6},\nonumber\\
	G^{(3)}(x_j^t, x_k^t) &=  \dfrac{1-\cos(x_j^t+x_k^t)}{2}.
\end{align} 
This oscillator dynamics is the optoelectronic dynamics defined in Ref.\cite{cho2017stable} with added triadic interactions. We also use this dynamics in \cref{subsec: stab}. Here, we chose the parameters $\alpha=0.5$ and $\sigma^{(3)}=1.5$. \cref{fig: clusters}(c) demonstrates the dynamics of two clusters (teal and violet) on the hypergraph shown in \cref{fig: clusters}(a), and \cref{fig: clusters}(f) demonstrates the dynamics of these clusters on the hypergraph shown in \cref{fig: clusters}(d). The dynamics of the two-cluster state are clearly distinct for these different hypergraph topologies with the same projected network.

Here, we covered one of the mechanisms that leads to non-isomorphic hypergraphs having the same projected adjacency matrix. Namely, it requires picking two isomorphic hypergraphs, and breaking the isomorphism by adding the same set of additional hyperedges. Clearly, if additional identical interactions of any order are present in both hypergraphs, \cref{eq: distinct i} still holds and the hypergraphs will have the same projection.

Note that in case of complete synchronization, distinct hypergraphs with the same dyadic projection produce the same effective behavior, so this phenomenon only arises for more complicated dynamical states.

\subsection{Does loss of information from the projection occur frequently in randomly selected hypergraphs?}\label{subsec: proj occurence}

To estimate if the information loss from projecting the hypergraph is frequent for a given number of nodes ($n_\text{nodes}$) and hyperedges ($n_\text{edges}$), we investigate how often the condition in \cref{eq: distinct i} holds for pairs of hypergraphs with hyperedges added at random.

\begin{figure}[]
	\includegraphics[scale=.55]{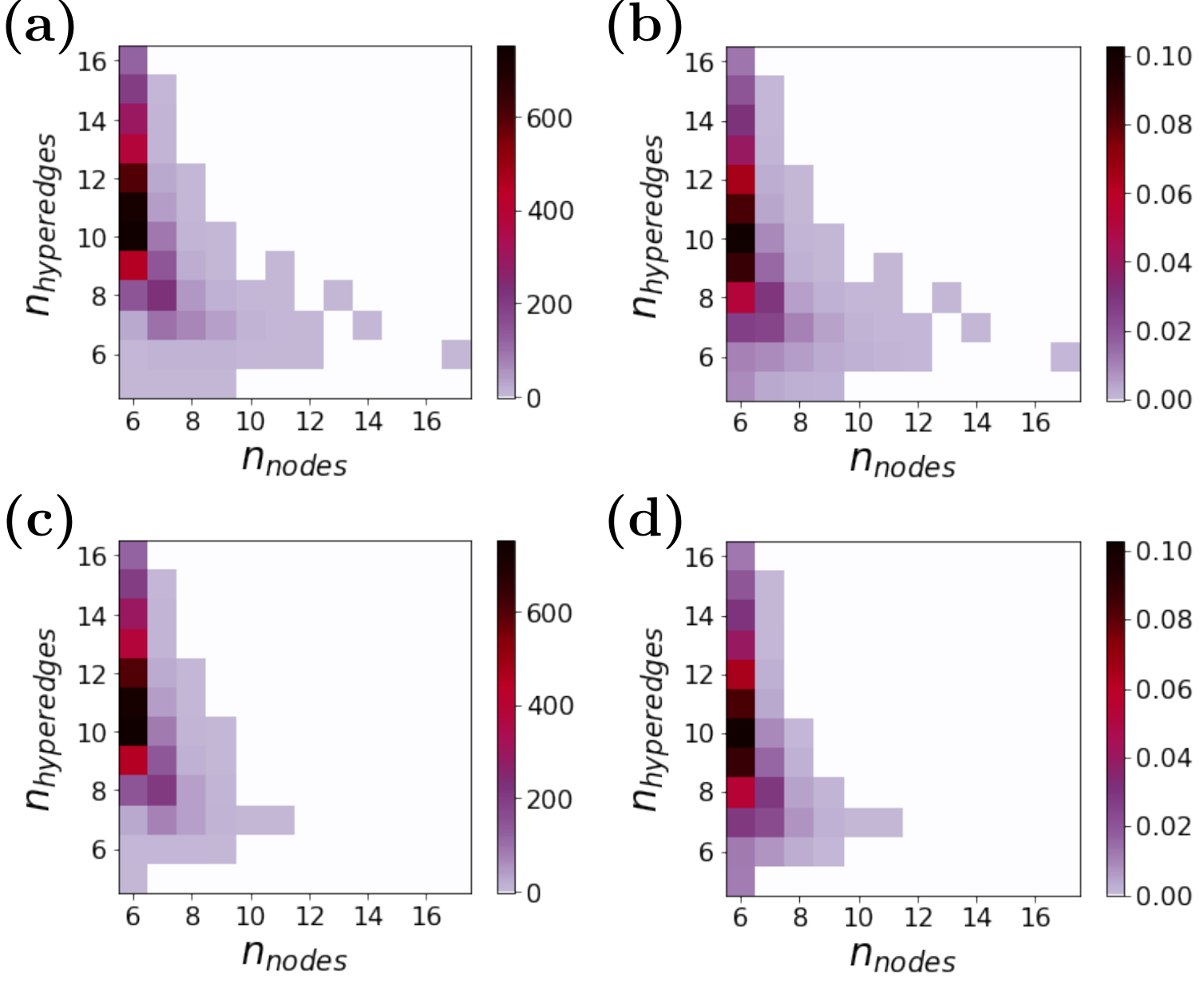}
	\caption{(a) Number and (b) fraction of pairs of non-isomorphic hypergraphs sharing the same projection for a given number of nodes and triadic hyperedges. 
		(c) Number and (d) fraction of pairs of non-isomorphic fully connected hypergraphs sharing the same projection. 
	}
	\label{fig: proj stats}
\end{figure}

It is known that bipartite network projections may exhibit data loss. In fact, it was shown that in some cases, non-isomorphic bipartite networks with incidence matrices $I_1$ and $I_2$ can have identical projections corresponding to node and edge interactions, i.e., $I_1 I_1^T= I_2 I_2^T$ and $I_1^T I_1= I_2^T I_2$, but those cases are rare \cite{kirkland2018two}. Here, we consider a similar problem in the context of hypergraphs, but only require the node interaction projections to be identical. In fact, for the example discussed in \cref{fig: clusters}, $I_1^T I_1 \neq I_2^T I_2$. This leaves us a wider range of options to explore. On the other hand, since we consider hypergraph projections where different edge orders can be distinguished, we only focus on matrices $I$ with constant column sums, where these sums equal to the edge order. This restriction narrows down the types of incidence matrices we consider.

Here, we focus on the case of triadic interactions on hypergraphs. First, we create $10,000$ random hypergraphs by randomly selecting with replacement three distinct nodes that will be connected by a hyperedge $n_\text{edges}$ times for each of the hypergraphs consisting of $n_\text{nodes}$ nodes. We note that the resulting hypergraphs may have duplicate edges, isolated nodes, or several connected components. We accept this since in real hypergraphs, more than one edge order can be present, so the nodes that are isolated for a specific interaction order may not be isolated when all orders of interactions are considered. But we also compare the results to those considering only fully connected hypergraphs. Then, we remove the ``duplicate'' isomorphic hypergraphs. Finally, we find the non-isomorphic hypergraphs satisfying \cref{eq: distinct i} and calculate the number and fractions of such pairs for each number of nodes and hyperedges. The results calculated for small numbers of nodes and hyperedges are presented in \cref{fig: proj stats}. We note that while these hypergraphs are not common, they could still occur as motifs in larger hypergraphs. For example, consider two identical hypergraphs. Adding different extra hyperedges to the same subset of their nodes, s.t. those hyperedges satisfy \cref{eq: distinct i} makes the whole hypergraph satisfy \cref{eq: distinct i}, producing two hypergraphs that are not isomorphic but have the same projection.

\section{Symmetry differences between hypergraphs and their dyadic projections}\label{sec: symm}

Structural symmetries of hypergraphs and dyadic networks determine some of the types of synchronization patterns admissible in the system and assist in determining their stability. We demonstrate that in some cases, there are symmetries of the projected adjacency matrix that are not the symmetries of the original hypergraph. 

Specifically, consider the permutations of each order of interactions defined in Ref.\cite{gambuzza2020master}
\begin{align}
P\mathcal{L}^{(m)}=\mathcal{L}^{(m)}P,
\end{align}
or, equivalently,
\begin{align}\label{eq: wrong symm}
P\mathcal{A}^{(m)}=\mathcal{A}^{(m)}P,
\end{align} 
where the permutation matrices $P$ satisfying \cref{eq: wrong symm}  for each order of interactions form the symmetry group. For the examples studied in Ref.\cite{gambuzza2020master}, the resulting symmetry group is associated with cluster synchronization states on simplicial complexes. We demonstrate that is not always the case, both in case of dynamics on simplicial complexes, or, more generally, hypergraphs.  
While this issue does not arise very often in randomly selected large hypergraphs, it is still important to know that using the conditions in Ref. \cite{gambuzza2020master} to obtain the patterns of cluster synchronization
requires an extra step of checking that each specific pattern is admissible on the full hypergraph and not just the projections of every order. 
Thus, next in \cref{subsec: symm}, we develop the conditions for cluster synchronization to arise from symmetries that hold for any hypergraph structure.

\subsection{Hypergraph symmetries}\label{subsec: symm}

Equitable partitions (groupings of nodes into clusters, in which nodes in the same cluster receive the same input from that cluster as well as all the other clusters) give rise to the admissible cluster synchronization states for a given hypergraph structure.
Equitable partitions that result from structural symmetries of the hypergraph are called orbital partitions and are a special case of more general equitable partitions \cite{golubitsky2003symmetry,golubitsky2012singularities}. For instance, all the partitions shown in \cref{fig: partitions} are orbital partitions, while the ones shown later in \cref{fig: equitable} are not. Even more flexibility is allowed for Laplacian-like coupling which requires only external equitable partitions  (groupings of nodes into clusters, in which nodes in the same cluster receive the same input from all the \textit{other} clusters, meaning that the hyperedges only containing one type of node cluster, e.g., the edge $e_{123}$ in \cref{fig: partitions}(a), can be ignored for admissibility purposes), and patterns arising from symmetries are less common in that case \cite{salova2021h1}. In summary, orbital partitions are a subset of equitable partitions, which are the subset of external equitable partitions.

Our focus in this section is structural symmetries. First, we state the algorithm for finding symmetry induced cluster synchronization patterns in systems with dyadic interactions. The automorphism group of the dyadic adjacency matrix $A$ is formed by a set of permutation matrices $P$, s.t. $PA=AP$. Any subgroup of that group can be linked to an admissible cluster synchronization pattern via orbital partitions. Namely, all the subsets of the network nodes that get mapped to themselves (and thus belong to the same cell of the orbital partition) can be completely synchronized \cite{pecora2014cluster}. The approach can be generalized to systems with different types of nodes and interactions, e.g., multilayer networks of coupled oscillators where cluster synchronization requires compatibility between intra- and interlayer symmetries \cite{della2020symmetries}. 

Symmetries of dyadic projected networks can not be immediately translated to those of a system with higher order interactions similarly to more general equitable partition methods. 
Instead, one has to consider the full hypergraph and the node and edge permutations simultaneously to assess the hypergraph synchronization patterns from the symmetry perspective. Symmetries, such as the ones analyzed in Ref.\cite{mulas2020hypergraph} for directed hypergraphs,  arise from the hypergraph automorphism group with elements $P$ represented as permutation matrices. We formulate the cluster synchronization condition in terms of the symmetries of the undirected hyperedges of each order $m$ as: 
\begin{align}\label{eq: commute}
	PI = I P_{\text{edge}}.
\end{align}
Here, $P_{N\times N}$ is a permutation matrix that reorders the nodes, and $[P_{\text{edge}}]_{M\times M}$ corresponds to the permutations of the edge labels if node labels are permuted.
These hyperedge permutation matrices are defined as follows:
\begin{align}\label{eq: edge permutation}
	[P_{\text{edge}}]_{e_i,e_j}=[P]_{i_1 j_1}...[P]_{i_m j_m},
\end{align} 
where $e_i=\{i_1,...,i_m\}$ and $e_j=\{j_1,...,j_m\}$ are the hyperedges.
The orbits of the subgroups of the automorphism group with elements $P$ determine the
admissible cluster synchronization patterns. 

Note that $I$ here is an aggregate matrix combining all the interaction orders. Alternatively, we could consider the incidence matrices for different orders of interactions, $I^{(m)}$, separately. 
Then, the largest common subgroup of the symmetry groups of all the interaction orders determines the automorphism group of the hypergraph.

As an example, consider the incidence  matrix corresponding to the hypergraph structure in \cref{fig: partitions}
\begin{align}
	I^{(3)}= \begin{pNiceMatrix}[first-row,first-col,nullify-dots]
		&\rotatebox{90}{[$123$]}&\rotatebox{90}{[$145$]}&\rotatebox{90}{[$246$]}&\rotatebox{90}{[$356$]}\\
		1 &{1}&1& & \\
		2 &1& &1& \\
		3 &1& & &1\\
		4 & &1&1& \\
		5 & &1& &1 \\
		6 & & &1&1\\
	\end{pNiceMatrix}.
\end{align}
One of the pairs containing a node permutation and its induced edge permutation satisfying \cref{eq: commute} is
\begin{align}
	P&=(1)(6)(2,3)(4,5),\nonumber\\
	P_{\text{edge}}&=([1,2,3],[1,4,5])([2,4,6],[3,5,6]).
\end{align}
Permuting the nodes and edges simultaneously leaves the structure of the hypergraph in \cref{fig: partitions} invariant.

\begin{figure*}
	\includegraphics[scale=.45]{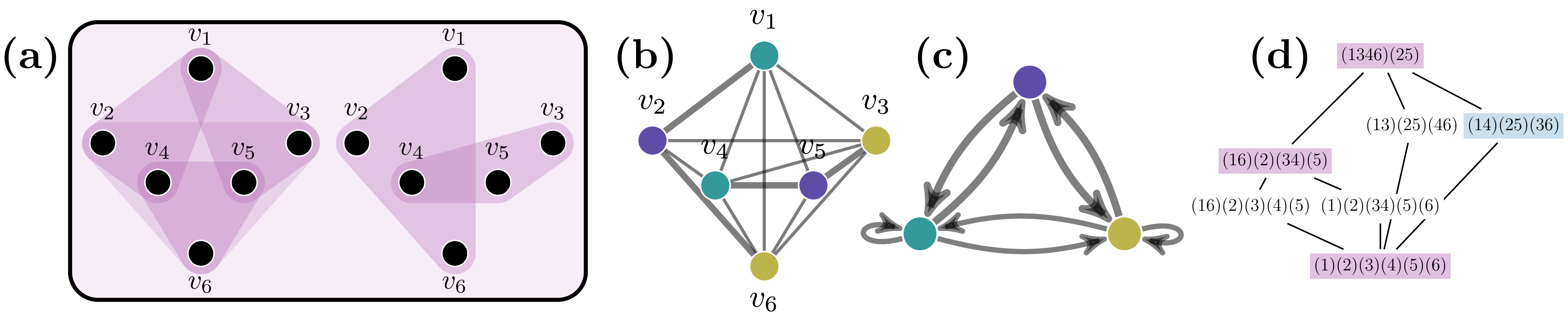}
	\caption{Symmetries of the dyadic projection (b) identify some patterns that are not admissible on the original hypergraph (a). (a) Hypergraph structure. (b) Hypergraph projection with a pattern of synchronization not admissible for the original hypergraph. (c) Quotient hypergraph of the projected network. (d) Lattice of partitions representing the admissible synchronization patterns obtained from the projection in (b). Only patterns highlighted in violet are admissible for the original hypergraph. Blue: pattern on [(b-c)].   
	}
	\label{fig: symm}
\end{figure*}

\subsection{Symmetries and square projection matrices}
Projected adjacency (or, if appropriate, Laplacian) matrices are useful to study cluster synchronization from the symmetry perspective \cite{gambuzza2020master}. However, in some cases, synchronization patterns obtained from the projected matrix are not admissible as synchronization patterns of the original hypergraph.

In \cref{subsec: symm}, we demonstrated how the symmetries of a hypergraph can be obtained from the incidence matrix. Equivalently, such symmetries can be deduced from the adjacency matrix $\mathcal{M}$ of the hypergraph's bipartite representation with the additional requirement that the permutations are of the form where nodes are permuted with nodes, and edges are permuted with edges. 
Similarly to \cref{subsec: structure}, we add a diagonal matrix $\mathcal{R}$ to ensure that. The conditions are then
\begin{align}\label{eq: condition1}
	P_{\mathcal{M}}(\mathcal{M}+\mathcal{R})=(\mathcal{M}+\mathcal{R})P_{\mathcal{M}},
\end{align}
where $P_{\mathcal M}$ is an $(N+M)\times(N+M)$-dimensional permutation matrix. Just like in the case of other balanced equivalence relations, only the coarsest edge partitions for each node partition would be the ones that actually correspond to the hyperedge permutations. However, all the partitions of the nodes themselves are valid.

If instead of the full hypergraph we consider the projection matrix, its symmetries (elements of the automorphism group) satisfy the condition
\begin{align}\label{eq: condition2}
	P_{\mathcal{A}}^{(m)}\mathcal{A}^{(m)}=\mathcal{A}^{(m)}P^{(m)}_{\mathcal{A}},
\end{align}
for each order $m$.
Here, the permutation matrix $P_{\mathcal{A}}$ is an $N\times N$-dimensional permutation matrix.

\begin{figure}[]
	\includegraphics[scale=.55]{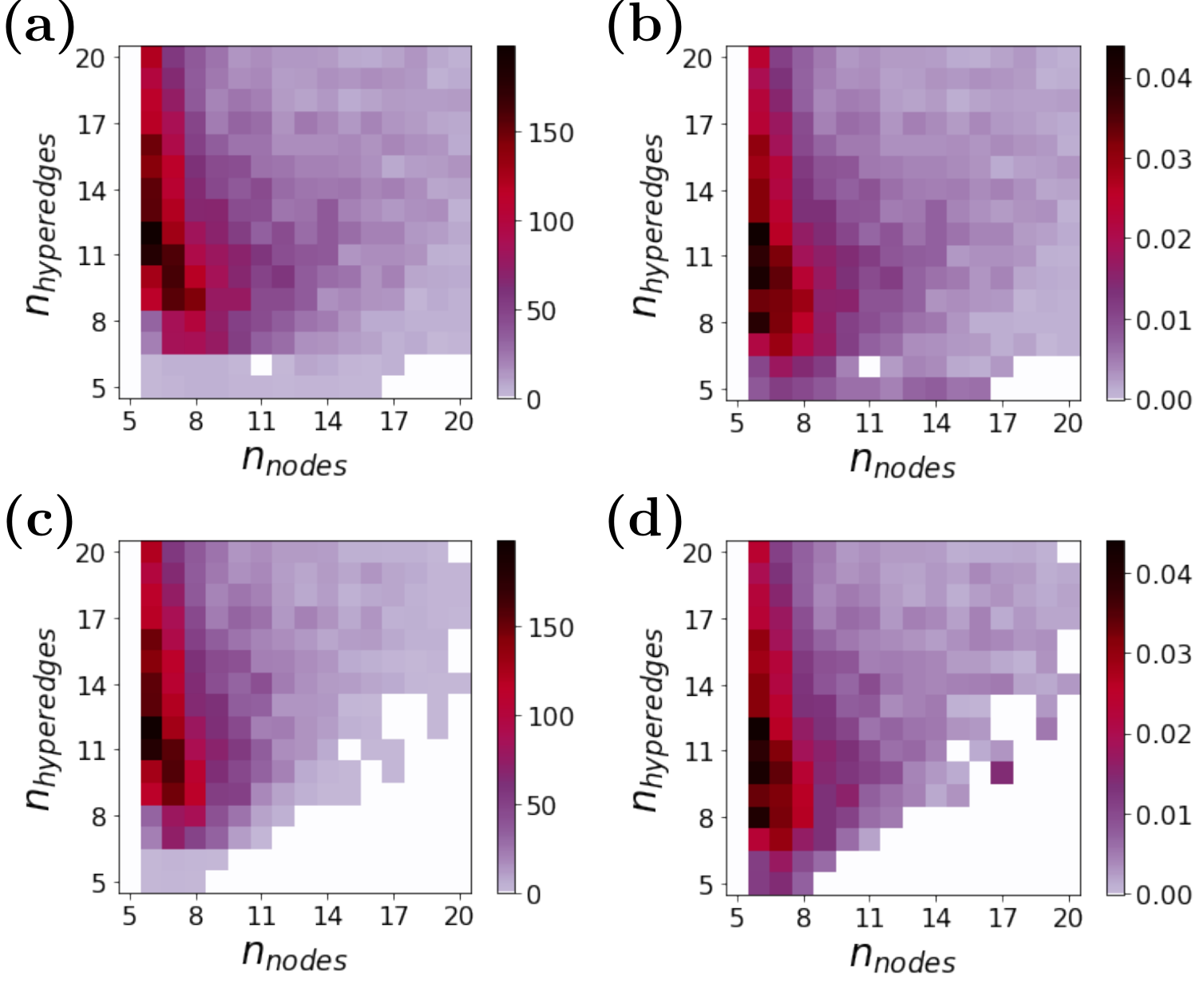}
	\caption{(a) Number and (b) fraction of hypergraphs with an automorphism group containing less elements than that of its projected network. (c) Number and (d) fraction of non-isomorphic fully connected hypergraphs with an automorphism group containing less elements than that of its projected network. }
	\label{fig: symm stats}
\end{figure}

First, we note that the node partitions for the original hypergraph obtained using \cref{eq: condition1} are always a subset of those for the projection shown in \cref{eq: condition2}.
This means that the admissible patterns on the hypergraph can be a proper subset of those that are admissible on the dyadic projection.
In many specific hypergraph cases, the conditions in \cref{eq: condition1} and \cref{eq: condition2} result in equivalent group orbits, thus corresponding to identical sets of admissible cluster synchronization patterns. However, in \cref{fig: symm}, we present an example when this equivalence does not hold. Specifically, \cref{fig: symm}(a) shows a hypergraph with six nodes and six hyperedges. This hypergraph admits three patterns of synchronization corresponding to the orbital partitions that are shaded in violet in \cref{fig: symm}(d). \cref{fig: symm}(b) demonstrates the structure of the dyadic projection of the hypergraph. This projection admits the full seven orbital partitions shown in \cref{fig: symm}(d).
One particular partition shaded in blue in \cref{fig: symm}(d) is specifically illustrated
in \cref{fig: symm}(b), where node colors correspond to a cluster assignment arising from \cref{eq: condition2}. \cref{fig: symm}(c) demonstrates the quotient network corresponding to the blue partition. Note that the blue partition does not correspond to a valid pattern of synchronization on the hypergraph in \cref{fig: symm}(a) since it does not satisfy \cref{eq: condition1}.

\subsection{How often are the symmetries of the hypergraph distinct from those of the projected network?}

We aim to get an idea of how often the automorphism group of the hypergraph does not have the same number of elements as that of its dyadic projection, which would result in different admissible cluster synchronization patterns. To do so, we generate a set of random non-isomorphic hypergraphs, similarly to \cref{subsec: proj occurence}. We then find the number of node partitions induced by the symmetry group of the bipartite network representing the hypergraph and compare that to the number of partitions induced by the symmetry group of the projected adjacency matrix, counting the number of cases in which these numbers are not the same.

Our results are presented in \cref{fig: symm stats}, where subfigures (a)-(b) consider all the hypergraphs we generate, and (c)-(d) only take into account hypergraphs with one connected component. 
While the number of occurrences when the hypergraph and the projection differ becomes relatively rare as the size of the hypergraph increases, the fact that they can differ means that if using, for instance, the method from Ref.\cite{gambuzza2020master}, an extra step of checking which of the orbital partitions of the projected hypergraph are the orbital partitions of the original hypergraph is required.

\begin{figure}[]
	\includegraphics[scale=.85]{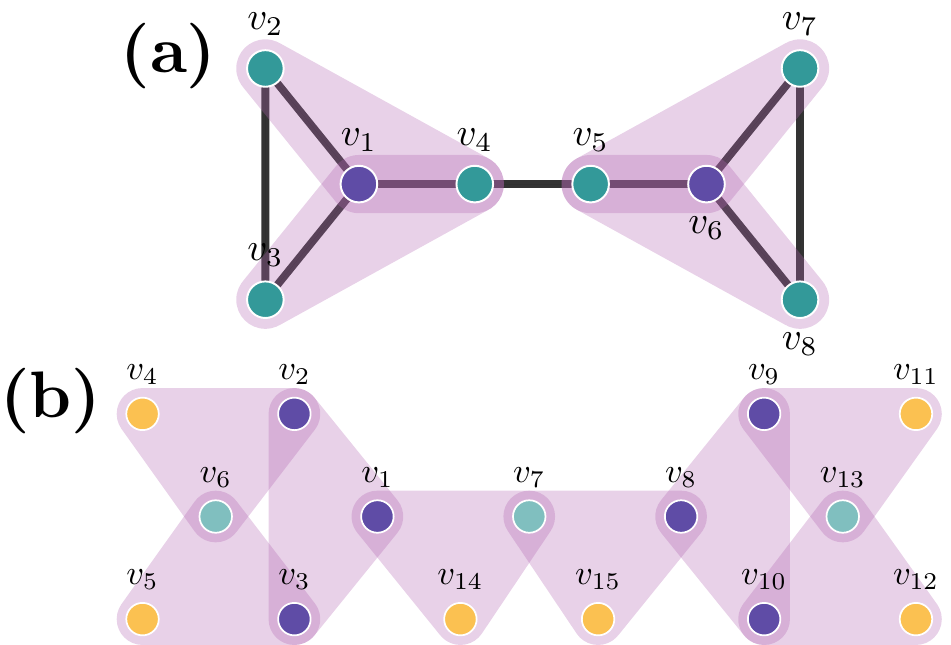}
	\caption{Examples of cluster synchronization patterns arising from equitable partitions that are not orbital partitions (in other words, patterns did not arise from symmetries). (a) Hypergraph with dyadic and triadic interactions. (b) Hypergraph with only triadic interactions. 
	}
	\label{fig: equitable}
\end{figure}

\section{Mismatch between equitable partitions of the hypergraph and the dyadic projection}

\begin{figure}[]
	\includegraphics[scale=.8]{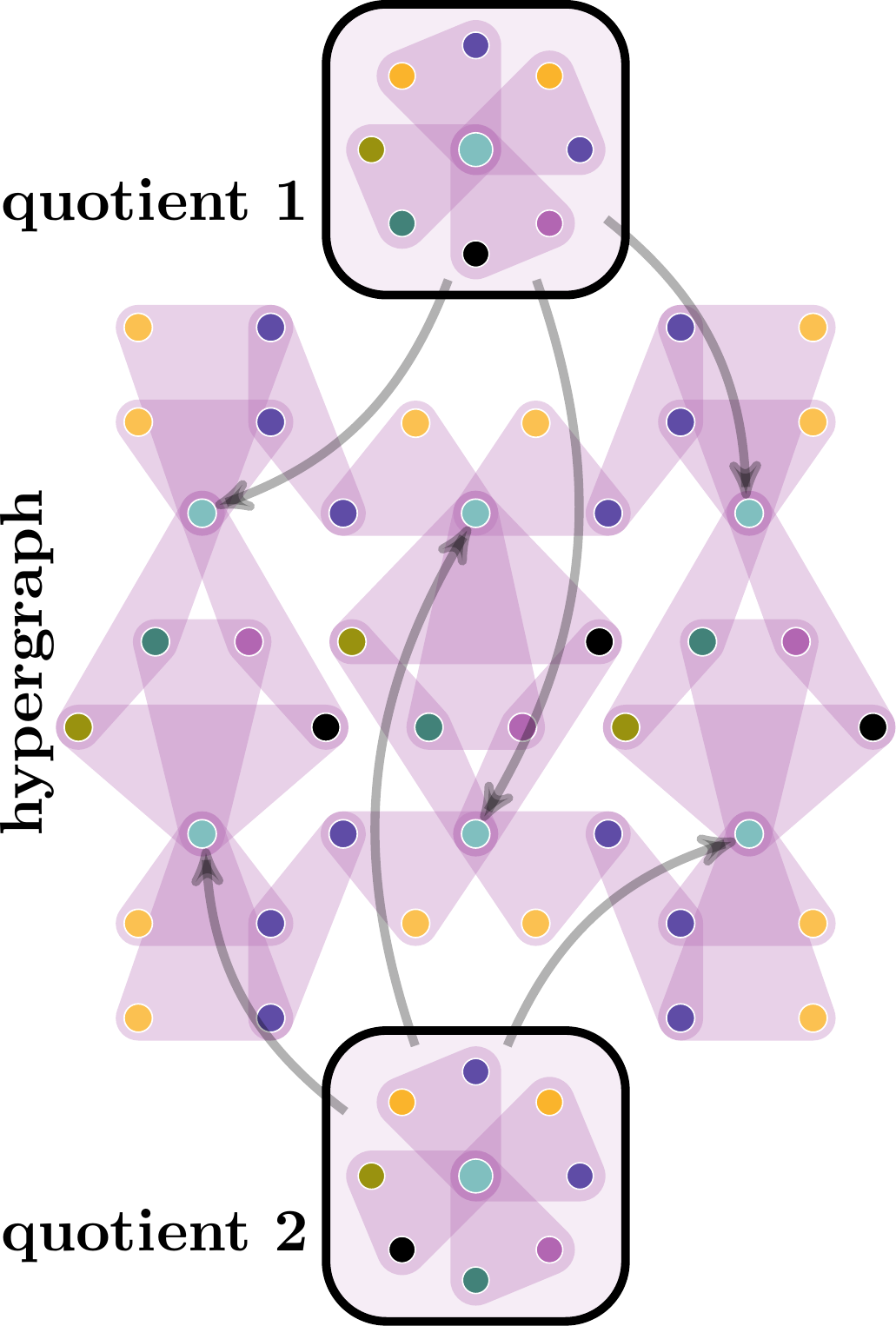}
	\caption{Example of a pattern that is not symmetry-induced, but follows from a more general equitable partition, which is not admissible on a hypergraph but is admissible on its projection. Top and bottom sub-figures (boxed, labeled as \textit{quotient 1} and \textit{quotient 2}) demonstrate the effective interactions of the nodes the arrows are pointing to. Since the effective interactions are distinct, the pattern is not admissible.}
	\label{fig: eq part projection}
\end{figure}

Beyond symmetries, more generally, admissible cluster synchronization patterns arise from equitable partitions (e.g., consider the cluster synchronization patterns in \cref{fig: equitable}(a,b)). One of the natural mechanisms for the latter is Laplacian and Laplacian-like coupling, where the fully synchronized edges do not add any dynamical contributions to the states of their nodes, and only affect the system's stability \cite{salova2021h1}. However, even for systems with non-diffusive (adjacency) coupling, clusters do not necessarily arise from the symmetries alone. Two such examples are shown in \cref{fig: equitable}. Here, \cref{fig: equitable}(a) is an example where the dyadic synchronization pattern does not arise from symmetries, with extra hyperedges added to form a system with higher order interactions. This can hold for systems with no dyadic interactions as well, as shown by the synchronization pattern in \cref{fig: equitable}(b).

If a pattern of synchronization arises from an equitable partition that is not an orbital partition, similar mismatch between the states of the full hypergraph and its projection can be observed. For instance, \cref{fig: eq part projection} demonstrates a pattern which is not symmetry induced but arises from an equitable partition. This pattern of cluster synchronization is not admissible for the full hypergraph, but is admissible on its projection.

\begin{figure}[]
	\includegraphics[scale=.5]{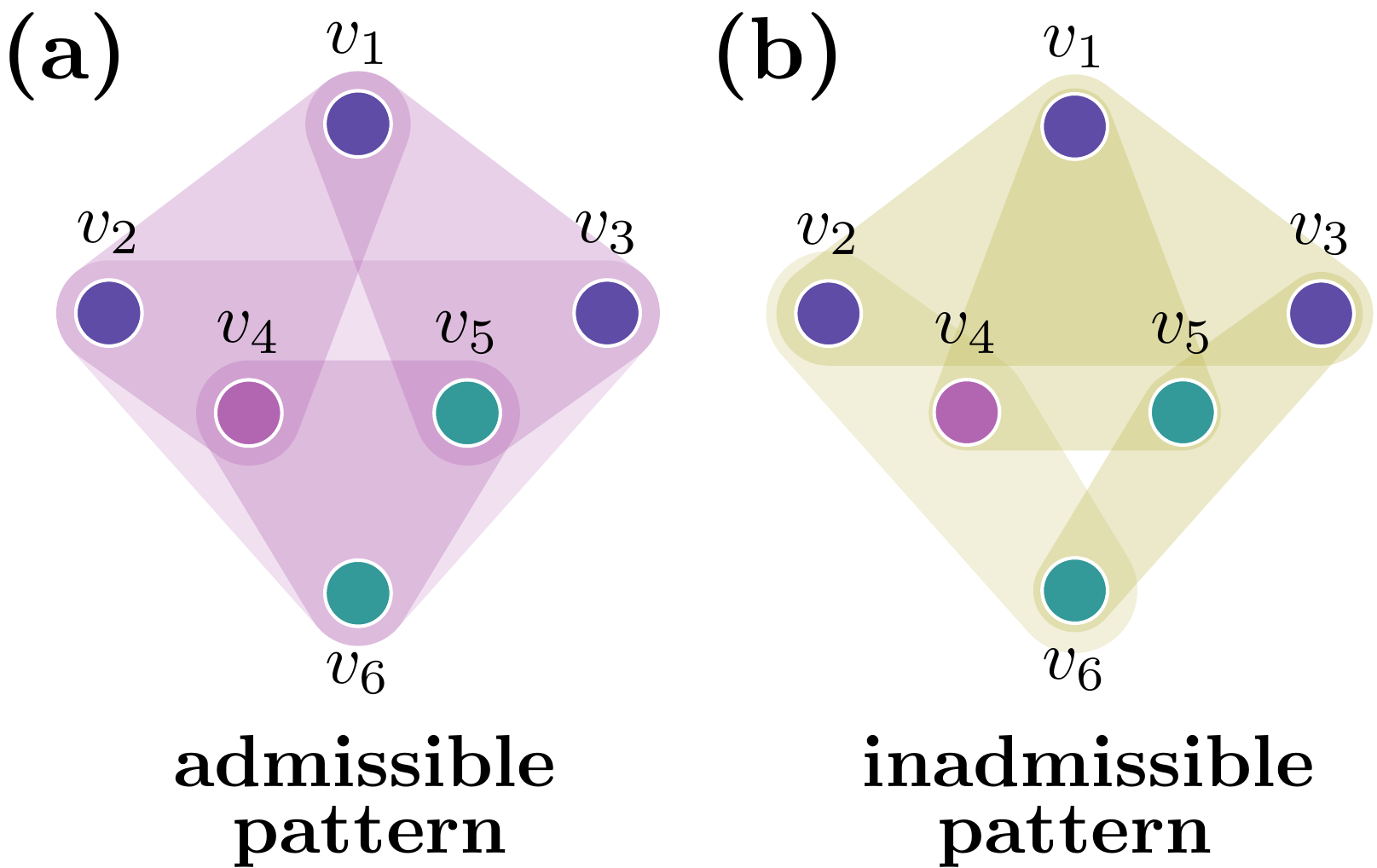}
	\caption{Three cluster pattern of synchronization (nodes in different clusters shown in different colors) on hypergraphs with the same dyadic projection. Assuming the coupling type defined in \cref{eq: ni}, this pattern of synchronization is admissible on the violet hypergraph (a), but not the olive hypergraph (b).}
	\label{fig: noninvasive}
\end{figure}

Additionally, for some types of coupling functions partitions more general than equitable partitions are sufficient, and it is also possible in that case that that the hypergraphs have the same projection but different admissible states. For instance, consider triadic coupling of the form
\begin{align}\label{eq: ni}
G(x_i,x_j,x_k)=g(x_j-x_i)g(x_k-x_i),
\end{align}
where $g(0)=0$, on a hypergraph with purely triadic interactions. The admissibility conditions for a cluster synchronization state given this coupling function are similar to \cref{eq: incidence cs}, with the caveat that a hyperedge influencing the node $i \in C_k$ can be ignored if any other node $j$ on a hyperedge is a part of the same cluster as $i$, i.e., $j \in C_k$, since then $G(x_i,x_j,x_k)=0$ when evaluated on that cluster synchronization state. Consider the isomorphic hypergraphs in \cref{fig: noninvasive}(a) (violet) and \cref{fig: noninvasive}(b) (olive). For the coupling form defined by \cref{eq: ni}, the same cluster assignment is admissible on \cref{fig: noninvasive}(a) (each node in a given cluster receives the same dynamical input), but not admissible on \cref{fig: noninvasive}(b) (nodes $1$, $2$, and $3$ are assigned to the same cluster, but node $3$ receives the dynamical input that is different from that received by nodes $1$ and $2$). Since the violet and olive hypergraphs have the same dyadic projection, dyadic projection alone would not be sufficient to determine which states are admissible for a given hypergraph structure. Therefore, having full information about the hypergraph structure is essential if the coupling functions allow us to relax some of the partition admissibility conditions.

\section{Stability calculations: Jacobian block diagonalization}\label{sec: stab}

\subsection{Background and dyadic interactions}
Here we consider the general case where equitable partitions and external equitable partitions determine admissible cluster synchronization patterns. 

Simultaneous block diagonalization, which can be based on symmetry considerations or balanced equivalence relations, is a useful tool that allows dimensionality reduction in cluster synchronization stability calculations. Tools for performing such reduction in the case of systems with dyadic interactions have been developed for systems such as networks with different types of edges, temporal networks, multilayer networks, and beyond.

Symmetry methods do not always provide the most refined block diagonal Jacobian structure \cite{zhang2020unified}. As shown in \cref{sec: symm}, while in most cases the symmetries of the hypergraph coincide with those of the projected matrix, this is not always the case. Regardless, once the symmetry group and group orbits are obtained, the irreducible representations of the symmetry group produce the block diagonalization of the Jacobian. The details of this process are discussed in \cref{app: symm}.

When  cluster synchronization partitions come from more general balanced equivalence relations, such as equitable or external equitable partitions, different methods must be used.
In systems with purely dyadic interactions, simultaneous block diagonalization of the coupling matrix and the diagonal indicator matrices corresponding to different clusters,
\begin{align}
	\{A, E_1,...,E_K\},
\end{align} 
block diagonalizes the Jacobian in a simple case of identical time-independent coupling. This dimensionality reduction can be performed using the algorithm from Refs. \cite{zhang2020symmetry,zhang2020unified}. Next, in \cref{subsec: stab}, we demonstrate how these results can be generalized to analyze cluster synchronization on hypergraphs.

\subsection{Stability of cluster synchronization on hypergraphs} \label{subsec: stab}

Here, we demonstrate how to perform and simplify symmetry-independent stability analysis for general undirected coupling (as opposed to Laplacian and Laplacian-like coupling discussed in Ref.\cite{salova2021h1}). The form of the dynamical equations and undirected coupling are discussed in \cref{eq: dynamics}. In this section, we develop cluster synchronization analysis for general patterns. The more specialized case of stability analysis of cluster synchronization patterns arising from symmetries is discussed in \cref{app: symm}.

As discussed in \cref{subsec: hyper proj}, a projected adjacency matrix for an interaction order $m$ can be defined as:
\begin{align}
	\mathcal{A}^{(m)}=I^{(m)}[I^{(m)}]^T-\mathcal D^{(m)},
\end{align}
where $[\mathcal D^{(m)}]_{ii}=\sum\limits_{j}I^{(m)}_{ij}$ and has zero off-diagonal elements. To analyze cluster synchronization on hypergraphs, instead of simply using $\mathcal{A}^{(m)}$, we need to define a projected adjacency matrix $\mathcal{A}^{(m)}_k$ for each interaction order $m$ and \textit{edge cluster} (distinct pattern of synchronization on a hyperedge of order $m$ induced by node clusters) indexed by $k$. Here,
\begin{align}\label{eq: Laplacian}
	\mathcal{A}^{(m)}_{k}=I^{(m)}_k[I^{(m)}_k]^T-\mathcal{D}^{(m)}_{k},
\end{align}
where $I^{(m)}_k$ is an $N\times |C^{(m)}_k|$ matrix (here, $|C^{(m)}_k|$ denotes the number of unique elements in the edge cluster $C^{(m)}_k$) consisting of the columns of $I^{(m)}$ corresponding to the hyperedges in the $k$th cluster of order $m$.  Additionally, $\mathcal D^{(m)}_{k}$ is a diagonal matrix of node degrees corresponding to the number of edges with that synchronization pattern ($[\mathcal D_k^{(m)}]_{ii}=\sum\limits_{j=1}^N [I^{(m)}_k]_{ij}$). 

Then, the variational equation determining the linear stability of cluster synchronization states can be expressed as:
\begin{align}\label{eq: stab}
	&{\delta \dot{x}}=
	\bigg(\sum\limits_{k=1}^K E_k \otimes JF(s_k)
	-\sum\limits_{m=2}^d\sigma^{(m)}\cdot\\
	&\Big[\sum\limits_{k=1}^{K_m}\sum\limits_{l\in \{C_k^{(m)}\}}\sum\limits_{p\in \{C_k^{(m)}\backslash l\}} E_{l} \mathcal{A}^{(m)}_kE_p\otimes JG^{(m)}(s_l,s_p,s_{C^{(m)}_k\backslash l,p})\nonumber\\  &+\sum\limits_{k=1}^{K_m}\sum\limits_{l\in \{C_k^{(m)}\}}E_l\mathcal{D}^{(m)}_k\otimes JG^{(m)}(s_l,s_{C^{(m)}_k\backslash l})\Big]\bigg)\delta x.\nonumber
\end{align}
Here, $\{C_k^{(m)}\}$ is the set of \textit{unique} node clusters included in the $k$th edge cluster. 
Additionally, $s_{C^{(m)}_k\backslash l}$ defines the set of all trajectories of nodes included in edge cluster $C^{(m)}_k$, excluding the one in the cluster $C_l$. Note that all node clusters and all edge clusters of all orders contribute to \cref{eq: stab}.

The terms contributing to the off-diagonal Jacobian elements are defined as
\begin{align}
	[JG^{(m)}&(s_l,s_p, s_{C^{(m)}_{k}\backslash l,p})]_{q,r} \\&=\dfrac{\partial G^{(m)}_q\left(x_i,x_{j},x_{k_1}...,x_{k_{m-2}}\right)}{\partial [x_{j}]_r}\bigg|_{\substack{x_i=s_l,x_j=s_p,\\x_{k_v}= [s_{C^{(m)}_k\backslash l,p}]_v}}.\nonumber
\end{align}
Similarly, the diagonal elements consist of 
\begin{align}\label{eq: derivs}
	[JF(s_k&)]_{q,r} =\dfrac{\partial F_q(x_i)}{\partial [x_i]_r}\bigg|_{x_i=s_k},\\ 
	[JG^{(m)}&(s_l,s_{C^{(m)}_{k}\backslash l})]_{q,r} \\&=\dfrac{\partial G^{(m)}_q\left(x_i,x_{k_1}...,x_{k_{m-1}}\right)}{\partial [x_{i}]_r}\bigg|_{\substack{x_i=s_l,\\x_{k_v}= [s_{C^{(m)}_k\backslash l}]_v}}.\nonumber
\end{align}
In addition, we note that if $i\in C_k$ and $j\in C_k$, $[\mathcal D_k^{(m)}]_{ii}=[\mathcal D_k^{(m)}]_{jj}$ from the cluster synchronization admissibility conditions.

While \cref{eq: stab} requires a lot of notation, the implication is simple: block diagonalization of the Jacobian requires the simultaneous block diagonalization of the set of cluster indicator matrices, the dyadic adjacency projection, and the projections for each higher order edge cluster:
\begin{align}\label{eq: mats hyper}
	\{E_1,...,E_K,\mathcal{A}^{(2)},\mathcal{A}^{(3)}_{1},...,\mathcal{A}^{(3)}_{K_3},...,\mathcal{A}^{(d)}_{1},...,\mathcal{A}^{(d)}_{K_d}\}.
\end{align}
The reason why only considering $\mathcal{A}^{(2)}$ for dyadic interactions is sufficient is discussed in \cref{subsec: app}.

For some topologies and synchronization patterns in systems with higher order interactions, it is only necessary to simultaneously block diagonalize the cluster indicator matrices and the projections of higher order coupling matrices, i.e.
\begin{align}\label{eq: proj bd}
	\{E_1,...,E_K, \mathcal A^{(2)},...,\mathcal A^{(d)}\}.
\end{align} 
 One of such cases, in addition to complete synchronization, is the case of noninvasive clusters \cite{zhang2020unified}, where each node in cluster $C_i$ receives the input from the same set of nodes in cluster $C_j$. Others include the case when there is only one edge pattern in the hypergraph for coupling of each order, i.e. there is only one matrix $\mathcal{A}^{(m)}_1=\mathcal{A}^{(m)}$ (e.g., \cref{fig: partitions}(c)), discussed in more detail in \cref{subsec: app}. 

For more general cases, such as the cluster synchronization patterns shown in \cref{fig: partitions}(a,b,d) and in \cref{fig: clusters,fig: equitable}, the conditions in \cref{eq: proj bd} are not sufficient. Thus, the direct analogy with dyadic coupling does not work, and the entire set of matrices in \cref{eq: mats hyper} is required for simultaneous block diagonalization.

\begin{figure*}
	\includegraphics[scale=.6]{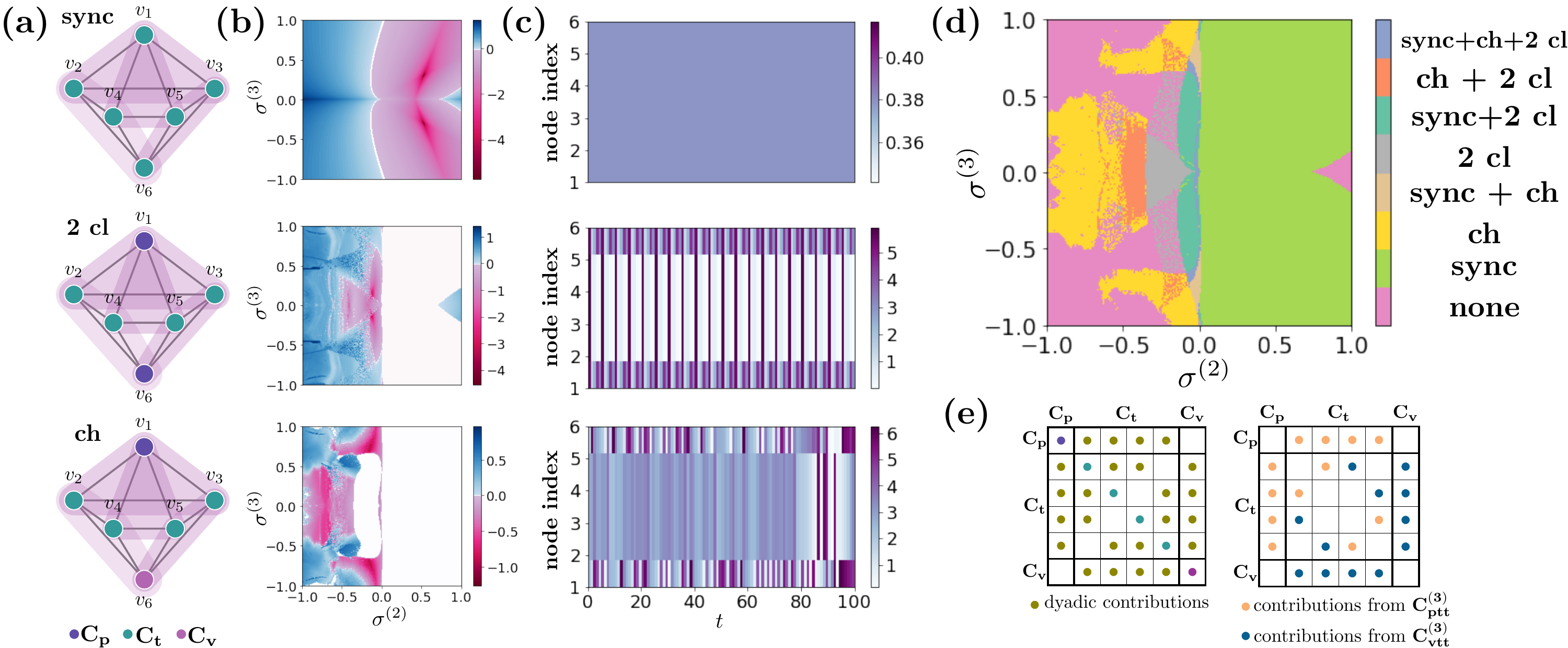}
	\caption{Stability of complete and cluster synchronization for a hypergraph of optoelectronic oscillators. (a) Synchronization patterns: full synchronization (top row), two clusters (middle row), and three-cluster chimera state (bottom row). (b) Linear stability (maximum transverse Lyapunov exponent) as a function of coupling strengths $\sigma^{(2)}$ and $\sigma^{(3)}$ for the states in part (a) based on $2.5\times 10^4$ time steps. White regions are due to not seeing a specific state in simulations on the quotient hypergraph for 20 initial conditions. (c) Representative trajectories for each state. Full synchronization state is a fixed point, 2-cluster state is periodic with period 5, 3-cluster state can be considered a chimera state and is chaotic. (d) Regions of stability of these states plotted together. It is evident there are regions where the system is multistable. (e) Matrices that need to be simultaneously block diagonalized for stability calculation. Distinct matrices shown in distinct colors. Diagonal elements correspond to the cluster indicator matrices $E_i$.}
	\label{fig: stab} 
\end{figure*}

To provide a concrete example, we consider the dynamics of a form:
\begin{align} \label{eq: concrete dynamics}
	x_{i}^{t+1}=F(x_i^t)+ \sum\limits_{m=2}^{d}\sum\limits_{e\in \mathcal{E}^{(m)}} [I^{(m)}]_{i,e} G^{(m)}(x_{e\backslash i}^t),
\end{align}
which is a discrete time analogue of  \cref{eq: dynamics}.
Additionally, we impose the optoelectronic oscillator dynamics as discussed in Ref.\cite{cho2017stable} but add higher order terms. Namely, we use 
\begin{align}
	F(x_i^t)&=\alpha\dfrac{1-\cos(x_i^t)}{2}+\dfrac{\pi}{6},\nonumber\\
	G^{(2)}(x_j^t) &= \sigma^{(2)} \dfrac{1-\cos(x_j^t)}{2},
\end{align}
where $\alpha = 2\pi/3-4\sigma^{(2)}$,
with additional higher order terms
\begin{align}
	G^{(3)}(x_j^t, x_k^t) &= \sigma^{(3)} \dfrac{1-\cos(x_j^t+x_k^t)}{2}
\end{align}
introduced to add triadic hyperedge interactions. Finally, we combine this dynamics with the hypergraph structure displayed in \cref{fig: stab}(a) to define our example system. 

The hypergraph in \cref{fig: stab}(a) consists of six coupled oscillators and supports a variety of cluster synchronization states: twenty two distinct cluster synchronization patterns are admissible, including the two extreme states where all the node trajectories are distinct and the fully synchronized state. To narrow down our analysis, we focus on three states: the fully synchronized, two-cluster, and three-cluster state as shown respectively in the rows of \cref{fig: stab}(a). The stability properties are shown in \cref{fig: stab}(b), and example trajectories in \cref{fig: stab}(c). The three-cluster state can be considered a chimera state in some regions of the phase space, since each cluster exhibits chaotic behavior that is not frequency synchronized with the other clusters (an example trajectory is shown in \cref{fig: stab}(c) bottom row). The set of matrices that need to be block diagonalized to analyze this state is demonstrated in \cref{fig: stab}(e). It is evident that higher order interactions have a large effect on stability, as stability regions change significantly when the triadic coupling $\sigma^{(3)}$ becomes nonzero. The numerically calculated stability regions of all three states are shown together in \cref{fig: stab}(d) as a function of $\sigma^{(2)}$ and $\sigma^{(3)}$.

\section{Conclusion}\label{sec: conclusion}

In this manuscript, we consider the applicability of dyadic methods for analyzing systems with higher order interactions in the context of cluster synchronization. Specifically, we consider the questions of admissibility and stability of the cluster synchronization states on hypergraphs. We show that the dyadic projection cannot be used in the most general instance and instead develop an analysis based on node and edge clusters which is sufficient.

First, we demonstrate that it is not always possible to reconstruct the hypergraph from its projected network. Since it is possible to construct the projections that distinguish between different orders of interactions, we focused our attention on distinct orders of interaction. While the cases when the information about the hypergraph is lost in the projections appear to be rare, they have strong implications on the analysis of cluster dynamics when distinct hypergraphs with identical projections both admit a specific cluster synchronization state, but have distinct quotient network structure dictating distinct dynamical evolution, as shown in \cref{fig: clusters}. 

Additionally, we investigate how the patterns of synchronization admissible on the hypergraph are related to those admissible on its dyadic projected network and demonstrate that these patterns do not have to be the same. We explicitly provide examples where some of the symmetries of the projected hypergraph do not preserve the structure of the hypergraph itself, making these patterns of synchronization inadmissible.

Dyadic methods still can be used to find the admissible patterns on a hypergraph by using the adjacency matrix of the corresponding bipartite graph with additional diagonal elements. However, for large hypergraphs, it may be more efficient to find the admissible patterns on the projected network and then manually check their admissibility on the original hypergraph.

Finally, while projected networks are sufficient to define the structure of the Jacobian for the stability calculation of the \textit{fully} synchronized state, it does not capture the full structure of the Jacobian for cluster synchronization. For networks with purely dyadic interactions, additional diagonal matrices describing the distinct node clusters are needed to capture the Jacobian structure. However, in case of higher order interactions, even that is not sufficient. We show that for higher order interactions we generally need multiple projection matrices, each corresponding to a specific order of interactions and specific \textit{edge} pattern of synchronization. These matrices are also useful in simplifying the stability calculations as shown in \cref{eq: mats hyper}.

The results on the admissibility of cluster synchronization obtained in this manuscript are easily generalizable to directed hypergraphs, just like the results for admissible clusters on dyadic networks generalize to networks with directed edges. The admissibility analysis we develop is valid for any cluster synchronization pattern on directed hypergraphs, provided that the edge clusters assignments take the hyperedge directedness into account.
Linear stability analysis for patterns arising from symmetries, as discussed in \cref{app: symm}, is also valid for directed hypergraphs.
The simplification of linear stability analysis and its interpretation for patterns arising beyond symmetry considerations can use the results from Refs.\cite{brady2021forget,lodi2021one}, which will again need to be generalized to higher order interactions.

In summary, hypergraph structures support a rich variety of dynamical phenomena, and hyperedges of all orders contribute to the dynamical evolution and stability calculations (see for instance \cref{eq: stab}). A formalism in terms of node clusters and edge clusters provides a principled way to organize the calculations for states beyond full synchronization. Such an approach may enable detailed analysis of the interplay between dyadic and higher order interactions and its impact on dynamical phenomena that can not be observed in systems with strictly dyadic interactions.

\vskip 1em
\appendix
\section{When projected matrices are enough for stability calculations}\label{subsec: app}
\begin{figure}[h]
	\includegraphics[scale=.58]{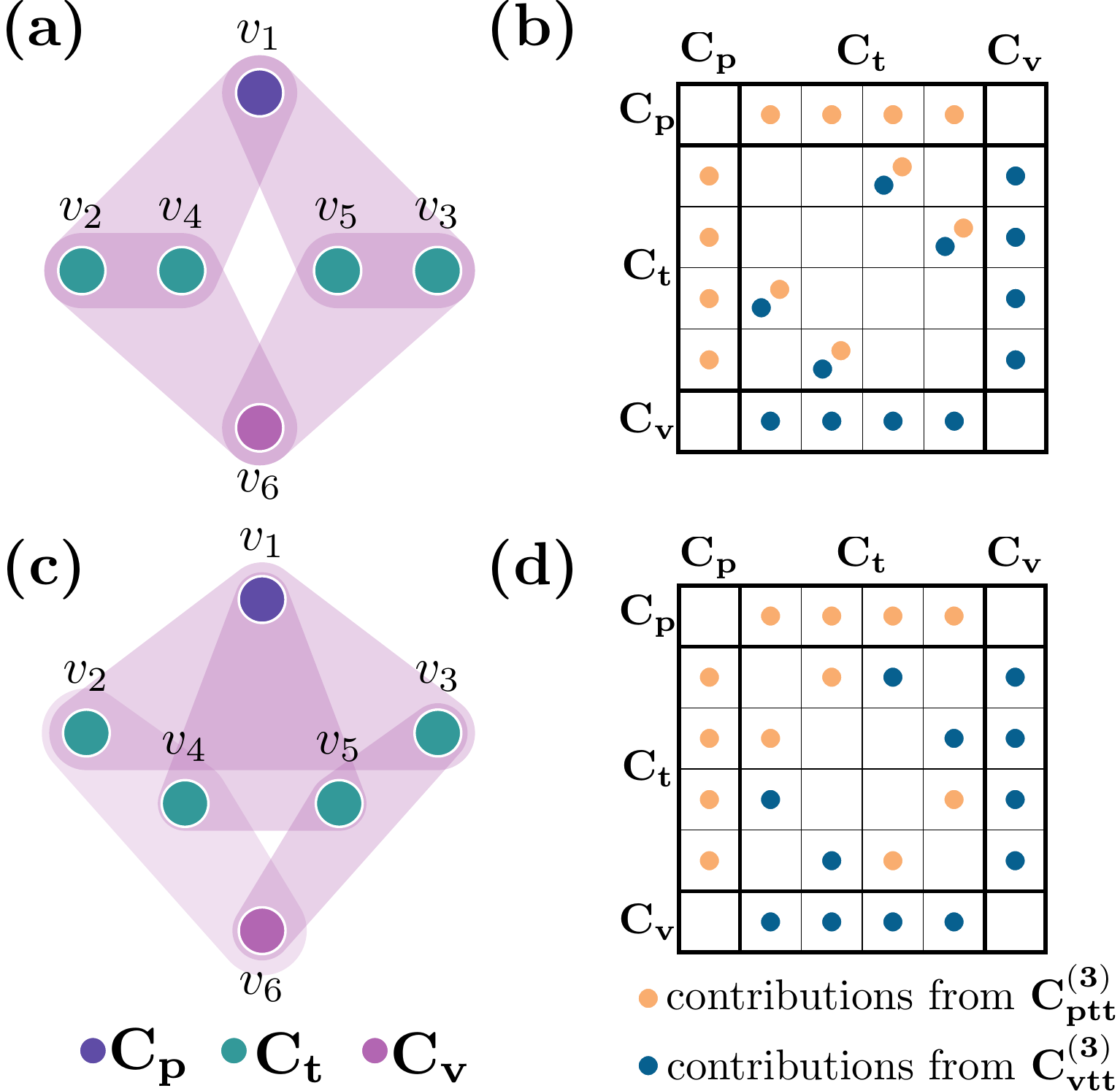}
	\caption{Example of a hypergraph and node clusters that it allows using \cref{eq: proj bd} for simultaneous block diagonalization. (a) Hypergraph and its node clusters, shown in distinct colors. (b) Jacobian structure. Distinct colors correspond to distinct edge contributions (e.g., orange corresponds to the derivatives of $G^{(3)}(s_1,s_2,s_3)$)}.
	\label{fig: simple j} 
\end{figure}

Here, we consider the cases in which the non-diagonal elements of each block of the Jacobian (i.e., each part corresponding to the interactions of specific clusters $C_i$ and $C_j$) coming from the same interaction order, denoted as $J^{(m)}_{C_i,C_j}$ contains only zeros and identical nonzero elements. 

The condition above clearly holds for systems with purely dyadic interactions, since each \textit{edge cluster} (e.g., connecting nodes in $C_i$ to nodes in $C_j$) only contributes to the Jacobian blocks $J^{(2)}_{C_i,C_j}$ and $J^{(2)}_{C_j,C_i}$.

The first way for the condition to hold for higher order edges is when for each pair of node clusters on a hyperedge, the edge cluster they are contained in is unique. In this case, the elements contributing to $J^{(m)}_{C_i,C_j}$ are either zero, or equal and unambiguously determined from \cref{eq: derivs}. For instance, consider the cluster synchronization pattern in \cref{fig: equitable}(a): the condition holds for its hyperedge clusters, $C^{(3)}_{ptt}$ and $C^{(3)}_{ppp}$. Similarly, it works for \cref{fig: equitable}(c): there is only one edge cluster, $C^{(3)}_{tvp}$. That condition, however, does not hold for \cref{fig: equitable}(b) and (d).

Now, assume there are several distinct edge clusters that contain the nodes from clusters $C_i$ and $C_j$, i.e., including a pair of edge clusters satisfying $\{i,j\}\subset C^{(m)}_k$ and $\{i,j\}\subset C^{(m)}_l$. Then, for every hyperedge in edge cluster $C^{(m)}_l$ containing the nodes $a\in C_i$ and $b\in C_j$, there has to exist a corresponding hyperedge in edge cluster $C^{(m)}_k$  that contains the nodes $a$ and $b$. This has to hold for every pair of such hyperedge clusters. The condition formulated here ensures that the nonzero off-diagonal Jacobian elements corresponding to the $m$th order interactions within each block are equal, and that they can be expressed as: 
\begin{align}
	J^{(m)}_{a,b} = \sum\limits_{k} JG^{(m)}(s_i,s_j, s_{C^{(m)}_{k}\backslash i,j}),
\end{align}
where $a\in C_i$, $b\in C_j$, and $\{i,j\}\subset C^{(m)}_k$.

As an example illustrating the discussion above, consider \cref{fig: simple j}(a) with three node clusters, $C_p$, $C_t$, and $C_v$, and two edge clusters, $C_{ttp}^{(3)}$ and $C_{ttv}^{(3)}$. There are two types of edges that contain two teal nodes, but both pairs ($2$ and $4$; $3$ and $5$) are contained exactly once in each type of hyperedges present in the hypergraph ($C_{ttp}^{(3)}$ and $C_{ttv}^{(3)}$), thus satisfying the requirements in the paragraph above. As illustrated in \cref{fig: simple j}(b), each block $J_{C_a,C_b}$ contains the same contributions within its nonzero elements, thus, the full adjacency matrix projections (\cref{eq: proj bd}) can be used for simultaneous block diagonalization. In contrast, consider \cref{fig: simple j}(c) with a hypergraph that supports the exact same pattern of synchronization but requires more intricate stability calculations. Block diagonalizing the Jacobian for \cref{fig: simple j}(c), the structure of which is schematically illustrated in \cref{fig: simple j}(d), requires a set of matrices from \cref{eq: mats hyper}, and exhibits different stability properties.

In summary, while under some circumstances block diagonalizing the matrices in \cref{eq: proj bd}  is sufficient, \cref{eq: mats hyper} is required in the most general case.

\section{Stability calculations for patterns arising from orbital partitions}\label{app: symm}

In the manuscript we discussed how some of the cluster synchronization patterns can be determined from symmetries (orbital partitions).
Due to a general result from equivariant dynamical systems theory, the Jacobian evaluated on a cluster synchronization state commutes with the elements of the symmetry group whose orbital partition determines the structure of that cluster synchronization state \cite{golubitsky2003symmetry} (we denote the actions of these elements by $P$). Therefore, the following holds:
\begin{align}\label{eq: gs}
	\mathcal J_{\text{cs}} P=P\mathcal J_{\text{cs}}.
\end{align}
where $\mathcal J_{\text{cs}}$ is the full Jacobian of the system  ($\delta \dot{x} = \mathcal J_{\text{cs}} \delta x$) evaluated at a particular cluster synchronization state.
As a result, the Jacobian can be block diagonalized using the matrices that block diagonalize the symmetry group elements $P$.

The individual terms of the variational equation for linear stability, \cref{eq: stab}, can also be used to demonstrate why the full Jacobian commutes with the symmetry group action.
First, we note that the diagonal cluster indicator matrices $E_k$ commute with the permutations $P$ ($PE_k=E_kP$), since the permutations $P$ only permute the nodes within a specific cluster.
Additionally, for the same reason, the matrices $P$ commute with the diagonal matrices $\mathcal D_k^{(m)}$.
We can also define the matrices $E_k^{(m)}$ to be diagonal matrices, s.t. $[E_k^{(m)}]_{ii}=1$ if the $i$th hyperedge of order $m$ belongs to $C^{(m)}_k$, and $[E_k^{(m)}]_{ii}=0$ otherwise. Similarly, the permutations $P_{\text{edge}}$ (as defined in \cref{eq: edge permutation}) commute with the matrices $E_k^{(m)}$, as they only permute the edges within a specific edge cluster.

Finally, we use \cref{eq: commute} to show that the matrices $\mathcal{A}^{(m)}_{k}=I^{(m)}_k[I^{(m)}_k]^T-\mathcal{D}^{(m)}_{k}$ commute with the action of the symmetry group:
\begin{align}
	&PI^{(m)}_k[I^{(m)}_k]^T=PI^{(m)}E_k^{(m)}[I^{(m)}E_k^{(m)}]^T
	\\&=I^{(m)}P_{\text{edge}}E_k^{(m)}[I^{(m)}]^T=I^{(m)}E_k^{(m)}P_{\text{edge}}[I^{(m)}]^T=\nonumber\\
	&= I^{(m)}E_k^{(m)}[I^{(m)}P_{edge}^{T}]^T=I^{(m)}E_k^{(m)}[I^{(m)}]^TP,\nonumber
\end{align}
using that $P^T=P^{-1}$ is an element of the symmetry group that preserves the structure of the hypergraph and generates the orbital partition leading to the cluster synchronization pattern we consider. Therefore, all the terms of \cref{eq: stab} commute with the symmetry group elements $P$, and as a result, \cref{eq: gs} holds.

Group representation theory can be used to block diagonalize the Jacobian to simplify the stability calculations in a similar way they are used for systems with dyadic interactions \cite{pecora2014cluster,sorrentino2020group}. Alternatively, other simultaneous block diagonalization methods are applicable and can result in a finer block diagonal structure \cite{zhang2020symmetry}. 

The steps in symmetry-based block diagonalization may appear simpler than those discussed in \cref{sec: stab}. Additionally, they apply to both directed and undirected hypergraphs. However, the calculation of irreducible representations that are then used to find the transformation of $\mathcal J_{\text{cs}}$ into the block diagonal form is more computationally expensive \cite{zhang2020unified}. Moreover, the method is only applicable to systems where the state arises from symmetries, and not the larger class of systems with patterns of cluster synchronization arising from balanced equivalence relations.

\bibliographystyle{unsrt}
\bibliography{biblio_hypergraphs}

\begin{thebibliography}{10}

\bibitem{barrat2008dynamical}
Alain Barrat, Marc Barthelemy, and Alessandro Vespignani.
\newblock {\em Dynamical processes on complex networks}.
\newblock Cambridge university press, 2008.

\bibitem{danon2011networks}
Leon Danon, Ashley~P Ford, Thomas House, Chris~P Jewell, Matt~J Keeling,
  Gareth~O Roberts, Joshua~V Ross, and Matthew~C Vernon.
\newblock Networks and the epidemiology of infectious disease.
\newblock {\em Interdisciplinary perspectives on infectious diseases}, 2011,
  2011.

\bibitem{rhoden2012self}
Martin Rohden, Andreas Sorge, Marc Timme, and Dirk Witthaut.
\newblock Self-organized synchronization in decentralized power grids.
\newblock {\em Phys. Rev. Lett.}, 109:064101, Aug 2012.

\bibitem{porter2020nonlinearity+}
Mason~A Porter.
\newblock Nonlinearity+ networks: A 2020 vision.
\newblock In {\em Emerging Frontiers in Nonlinear Science}, pages 131--159.
  Springer, 2020.

\bibitem{curto2019relating}
Carina Curto and Katherine Morrison.
\newblock Relating network connectivity to dynamics: opportunities and
  challenges for theoretical neuroscience.
\newblock {\em Current opinion in neurobiology}, 58:11--20, 2019.

\bibitem{bick2021higherorder}
Christian Bick, Elizabeth Gross, Heather~A Harrington, and Michael~T Schaub.
\newblock What are higher-order networks?
\newblock {\em arXiv preprint arXiv:2104.11329}, 2021.

\bibitem{battiston2020networks}
Federico Battiston, Giulia Cencetti, Iacopo Iacopini, Vito Latora, Maxime
  Lucas, Alice Patania, Jean-Gabriel Young, and Giovanni Petri.
\newblock Networks beyond pairwise interactions: structure and dynamics.
\newblock {\em Physics Reports}, 2020.

\bibitem{skardal2019abrupt}
Per~Sebastian Skardal and Alex Arenas.
\newblock Abrupt desynchronization and extensive multistability in globally
  coupled oscillator simplexes.
\newblock {\em Physical Review Letters}, 122(24):248301, 2019.

\bibitem{skardal2020memory}
Per~Sebastian Skardal and Alex Arenas.
\newblock Memory selection and information switching in oscillator networks
  with higher-order interactions.
\newblock {\em Journal of Physics: Complexity}, 2020.

\bibitem{gambuzza2020master}
LV~Gambuzza, F~Di~Patti, L~Gallo, S~Lepri, M~Romance, R~Criado, M~Frasca,
  V~Latora, and S~Boccaletti.
\newblock Stability of synchronization in simplicial complexes.
\newblock {\em Nature Communications}, 12(1):1--13, 2021.

\bibitem{zhang2020unified}
Yuanzhao Zhang, Vito Latora, and Adilson~E Motter.
\newblock Unified treatment of dynamical processes on generalized networks:
  Higher-order, multilayer, and temporal interactions.
\newblock {\em arXiv preprint arXiv:2010.00613}, 2020.

\bibitem{xu2020bifurcation}
Can Xu, Xuebin Wang, and Per~Sebastian Skardal.
\newblock Bifurcation analysis and structural stability of simplicial
  oscillator populations.
\newblock {\em Physical Review Research}, 2(2):023281, 2020.

\bibitem{landry2020effect}
Nicholas~W Landry and Juan~G Restrepo.
\newblock The effect of heterogeneity on hypergraph contagion models.
\newblock {\em Chaos: An Interdisciplinary Journal of Nonlinear Science},
  30(10):103117, 2020.

\bibitem{lucas2020multi}
Maxime Lucas, Giulia Cencetti, and Federico Battiston.
\newblock Multiorder laplacian for synchronization in higher-order networks.
\newblock {\em Phys. Rev. Research}, 2:033410, Sep 2020.

\bibitem{skardal2020higher}
Per~Sebastian Skardal and Alex Arenas.
\newblock Higher order interactions in complex networks of phase oscillators
  promote abrupt synchronization switching.
\newblock {\em Communications Physics}, 3(1):1--6, 2020.

\bibitem{millan2020explosive}
Ana~P Mill{\'a}n, Joaqu{\'\i}n~J Torres, and Ginestra Bianconi.
\newblock Explosive higher-order {Kuramoto} dynamics on simplicial complexes.
\newblock {\em Physical Review Letters}, 124(21):218301, 2020.

\bibitem{ghorbanchian2021higher}
Reza Ghorbanchian, Juan~G Restrepo, Joaqu{\'\i}n~J Torres, and Ginestra
  Bianconi.
\newblock Higher-order simplicial synchronization of coupled topological
  signals.
\newblock {\em Communications Physics}, 4(1):1--13, 2021.

\bibitem{deville2021consensus}
Lee DeVille.
\newblock Consensus on simplicial complexes: Results on stability and
  synchronization.
\newblock {\em Chaos: An Interdisciplinary Journal of Nonlinear Science},
  31(2):023137, 2021.

\bibitem{carletti2020dynamical}
Timoteo Carletti, Duccio Fanelli, and Sara Nicoletti.
\newblock Dynamical systems on hypergraphs.
\newblock {\em Journal of Physics: Complexity}, 1(3):035006, aug 2020.

\bibitem{de2021phase}
Guilherme~Ferraz de~Arruda, Michele Tizzani, and Yamir Moreno.
\newblock Phase transitions and stability of dynamical processes on
  hypergraphs.
\newblock {\em Communications Physics}, 4(1):1--9, 2021.

\bibitem{mulas2020coupled}
Raffaella Mulas, Christian Kuehn, and J{\"u}rgen Jost.
\newblock Coupled dynamics on hypergraphs: Master stability of steady states
  and synchronization.
\newblock {\em arXiv preprint arXiv:2003.13775}, 2020.

\bibitem{golubitsky2003symmetry}
Martin Golubitsky and Ian Stewart.
\newblock {\em The symmetry perspective: from equilibrium to chaos in phase
  space and physical space}, volume 200.
\newblock Springer Science \& Business Media, 2003.

\bibitem{pecora2014cluster}
Louis~M Pecora, Francesco Sorrentino, Aaron~M Hagerstrom, Thomas~E Murphy, and
  Rajarshi Roy.
\newblock Cluster synchronization and isolated desynchronization in complex
  networks with symmetries.
\newblock {\em Nature Communications}, 5:4079, 2014.

\bibitem{mulas2020hypergraph}
Raffaella Mulas, Rub{\'e}n~J S{\'a}nchez-Garc{\'\i}a, and Ben~D MacArthur.
\newblock Hypergraph automorphisms.
\newblock {\em arXiv preprint arXiv:2010.01049}, 2020.

\bibitem{stewart2003symmetry}
Ian Stewart, Martin Golubitsky, and Marcus Pivato.
\newblock Symmetry groupoids and patterns of synchrony in coupled cell
  networks.
\newblock {\em SIAM Journal on Applied Dynamical Systems}, 2(4):609--646, 2003.

\bibitem{salova2021h1}
Anastasiya Salova and Raissa~M D'Souza.
\newblock Cluster synchronization on hypergraphs.
\newblock {\em arXiv preprint arXiv:2101.05464}, 2021.

\bibitem{salova2021code}
Anastasiya Salova.
\newblock Admissibility and stability of cluster synchronization patterns on
  hypergraphs.
\newblock \url{https://github.com/asalova/hypergraph-cluster-sync}.

\bibitem{zhang2020symmetry}
Yuanzhao Zhang and Adilson~E. Motter.
\newblock Symmetry-independent stability analysis of synchronization patterns.
\newblock {\em SIAM Review}, 62(4):817--836, 2020.

\bibitem{PhysRevResearch.2.023281}
Can Xu, Xuebin Wang, and Per~Sebastian Skardal.
\newblock Bifurcation analysis and structural stability of simplicial
  oscillator populations.
\newblock {\em Phys. Rev. Research}, 2:023281, Jun 2020.

\bibitem{belykh2008cluster}
Vladimir~N. Belykh, Grigory~V. Osipov, Valentin~S. Petrov, Johan A.~K. Suykens,
  and Joos Vandewalle.
\newblock Cluster synchronization in oscillatory networks.
\newblock {\em Chaos: An Interdisciplinary Journal of Nonlinear Science},
  18(3):037106, 2008.

\bibitem{pecora2017discovering}
Louis~M Pecora, Francesco Sorrentino, Aaron~M Hagerstrom, Thomas~E Murphy, and
  Rajarshi Roy.
\newblock Discovering, constructing, and analyzing synchronous clusters of
  oscillators in a complex network using symmetries.
\newblock In {\em Advances in Dynamics, Patterns, Cognition}, pages 145--160.
  Springer, 2017.

\bibitem{cho2017stable}
Young~Sul Cho, Takashi Nishikawa, and Adilson~E Motter.
\newblock Stable chimeras and independently synchronizable clusters.
\newblock {\em Physical Review Letters}, 119(8):084101, 2017.

\bibitem{salova2020decoupled}
Anastasiya Salova and Raissa~M. D'Souza.
\newblock Decoupled synchronized states in networks of linearly coupled limit
  cycle oscillators, 2020.

\bibitem{bohle2021coupled}
Tobias Böhle, Christian Kuehn, Raffaella Mulas, and Jürgen Jost.
\newblock Coupled hypergraph maps and chaotic cluster synchronization.
\newblock {\em arXiv preprint arXiv:2102.02272}, 2021.

\bibitem{hagberg2008exploring}
Aric Hagberg, Pieter Swart, and Daniel S~Chult.
\newblock Exploring network structure, dynamics, and function using {NetworkX}.
\newblock Technical report, Los Alamos National Lab.(LANL), Los Alamos, NM
  (United States), 2008.

\bibitem{kirkland2018two}
Steve Kirkland.
\newblock Two-mode networks exhibiting data loss.
\newblock {\em Journal of Complex Networks}, 6(2):297--316, 2018.

\bibitem{golubitsky2012singularities}
Martin Golubitsky, Ian Stewart, and David~G Schaeffer.
\newblock {\em Singularities and groups in bifurcation theory}, volume~2.
\newblock Springer Science \& Business Media, 2012.

\bibitem{della2020symmetries}
Fabio Della~Rossa, Louis Pecora, Karen Blaha, Afroza Shirin, Isaac Klickstein,
  and Francesco Sorrentino.
\newblock Symmetries and cluster synchronization in multilayer networks.
\newblock {\em Nature Communications}, 11(1):1--17, 2020.

\bibitem{brady2021forget}
Fiona~M Brady, Yuanzhao Zhang, and Adilson~E Motter.
\newblock Forget partitions: Cluster synchronization in directed networks
  generate hierarchies.
\newblock {\em arXiv preprint arXiv:2106.13220}, 2021.

\bibitem{lodi2021one}
Matteo Lodi, Francesco Sorrentino, and Marco Storace.
\newblock One-way dependent clusters and stability of cluster synchronization
  in directed networks.
\newblock {\em arXiv preprint arXiv:2106.06611}, 2021.

\bibitem{sorrentino2020group}
Francesco Sorrentino, Louis~M Pecora, and Ljiljana Trajkovic.
\newblock Group consensus in multilayer networks.
\newblock {\em IEEE Transactions on Network Science and Engineering}, 2020.

\end{thebibliography}

\end{document}